\documentclass[11pt]{article}

\usepackage[margin=0.8in]{geometry}

\usepackage{varioref}
\usepackage[usenames,svgnames,xcdraw,table]{xcolor}
\definecolor{DarkBlue}{rgb}{0.1,0.1,0.5}
\definecolor{DarkGreen}{rgb}{0.1,0.5,0.1}
\usepackage[backref=page]{hyperref}
\hypersetup{
     colorlinks   = true,
     linkcolor    = DarkBlue, % color of internal links
     urlcolor     = DarkBlue, % color of external links
	 citecolor    = DarkGreen % color of links to bibliography
}
\renewcommand*{\backref}[1]{}
\renewcommand*{\backrefalt}[4]{%
    \ifcase #1 (Not cited.)%
    \or        (Cited on page~#2)%
    \else      (Cited on pages~#2)%
    \fi}
\usepackage[capitalise,noabbrev]{cleveref}

\usepackage{appendix}

\usepackage{algorithmic}
\usepackage{algorithm}

\usepackage{graphicx,epsfig,amsmath,latexsym,amssymb,verbatim}%,revsymb}

\newcommand{\extra}[1]{}

\usepackage{amsmath,amssymb,amsfonts}
\usepackage{amsthm}
\usepackage{thmtools,thm-restate}

\newtheorem{theorem}{Theorem}

\newtheorem{definition}[theorem]{Definition}
\newtheorem{lemma}[theorem]{Lemma}

\newtheorem{claim}{Claim}

\def\squareforqed{\hbox{\rlap{$\sqcap$}$\sqcup$}}
\def\qed{\ifmmode\squareforqed\else{\unskip\nobreak\hfil
\penalty50\hskip1em\null\nobreak\hfil\squareforqed
\parfillskip=0pt\finalhyphendemerits=0\endgraf}\fi}
\def\endenv{\ifmmode\;\else{\unskip\nobreak\hfil
\penalty50\hskip1em\null\nobreak\hfil\;
\parfillskip=0pt\finalhyphendemerits=0\endgraf}\fi}

\renewenvironment{proof}{\noindent \textbf{{Proof~} }}{\qed\medskip}
\newenvironment{proof+}[1]{\noindent \textbf{{Proof #1~} }}{\qed\medskip}

\mathchardef\ordinarycolon\mathcode`\:
\mathcode`\:=\string"8000
\def\vcentcolon{\mathrel{\mathop\ordinarycolon}}
\begingroup \catcode`\:=\active
  \lowercase{\endgroup
  \let :\vcentcolon
  }

\newcommand{\nc}{\newcommand}

\DeclareMathOperator*{\argmin}{arg\,min}
\DeclareMathOperator*{\argmax}{arg\,max}

\nc{\barA}{\overline{A}}
\nc{\barB}{\overline{B}}
\nc{\barC}{\overline{C}}
\nc{\barD}{\overline{D}}
\nc{\barR}{\overline{R}}
\nc{\barX}{\overline{X}}
\nc{\barY}{\overline{Y}}
\nc{\barU}{\overline{U}}

\nc{\ALG}{\textsc{Alg}}

\newcommand{\EFone}{\textrm{EF1}}
\newcommand{\epsEF}{\varepsilon\textrm{-EF}}
\newcommand{\mms}{\textsc{Mms}}
\newcommand{\MMS}{\textsc{Mms}}
\newcommand{\shortcite}{\cite}
\newcommand{\citeauthor}{\cite}
\begin{document}

\title{{\bfseries Fair Division with a Secretive Agent}}
\author{Eshwar Ram Arunachaleswaran\thanks{Indian Institute of Science. {\tt eshwarram.arunachaleswaran@gmail.com}} \qquad Siddharth Barman\thanks{Indian Institute of Science. {\tt barman@iisc.ac.in}} \qquad Nidhi Rathi\thanks{Indian Institute of Science. {\tt nidhirathi@iisc.ac.in}}}

\date{}
\maketitle
\begin{abstract}
	We study classic fair-division problems in a partial information setting. This paper respectively addresses fair division of rent, cake, and indivisible goods among agents with cardinal preferences. We will show that, for all of these settings and under appropriate valuations, a fair (or an approximately fair) division among $n$ agents can be efficiently computed using only the valuations of $n-1$ agents. The $n$th (secretive) agent can make an arbitrary selection after the division has been proposed and, irrespective of her choice, the computed division will admit an overall fair allocation. 
	
	For the rent-division setting we prove that the (well-behaved) utilities of $n-1$ agents suffice to find a rent division among $n$ rooms such that, for every possible room selection of the secretive agent, there exists an allocation (of the remaining $n-1$ rooms among the $n-1$ agents) which ensures overall \emph{envy freeness} (fairness). We complement this existential result by developing a polynomial-time algorithm that finds such a fair rent division under  quasilinear utilities. 
	
	In this partial information setting, we also develop efficient algorithms to compute allocations that are \emph{envy-free up to one good} ($\EFone$) and \emph{$\varepsilon$-approximate envy free}. These two notions of fairness are applicable in the context of indivisible goods and divisible goods (cake cutting), respectively. This work also addresses fairness in terms of \emph{proportionality} and \emph{maximin shares}. Our key result here is an efficient algorithm that, even with a secretive agent, finds a $1/19$-approximate maximin fair allocation (of indivisible goods) under submodular valuations of the non-secretive agents. 
	
	One of the main technical contributions of this paper is the development of novel connections between different fair-division paradigms, e.g., we use our existential results for envy-free rent-division to develop an efficient $\EFone$ algorithm. 
	
\end{abstract}

\section{Introduction}
\label{section:introduction}

The theory of fair division addresses the fundamental problem of dividing resources/goods in a fair manner among agents with heterogeneous valuations, but equal entitlements \cite{moulin1986game}. Many recent results in algorithmic game theory address computational aspects of fair division, see, e.g., \cite{brandt2016handbook} and \cite{rothe2015economics} for excellent expositions. A central thread of research here is to design algorithms that complement nonconstructive existence results. This computational perspective has lead to notable algorithms, hardness results, and the development of novel solution concepts; in fact, a number of such notions and algorithms have been integrated into widely-used websites (e.g., Spliddit\footnote{\url{http://www.spliddit.org/}} and Adjusted Winner\footnote{\url{http://www.nyu.edu/projects/adjustedwinner/}}), which provide methods for fair division of resources, rent, and tasks. This works contributes to computational thinking in fair division and, in particular, studies algorithmic aspects of existence results which show that the preferences of only $n-1$ agents suffice to find a fair division among $n$ agents. 
%the emergent theme of

Such an existence result for the case of two agents and a divisible good (which is metaphorically represented as a cake) follows directly from the standard {divide-and-choose protocol}: considering the valuation of just the first agent we can partition the cake into two parts of equal value (for the first agent) and, then, the second agent can select her most preferred piece. This protocol leads to an \emph{envy-free} cake division, i.e., in the resulting allocation no agent has a strictly stronger preference for the other agent's piece. Envy freeness is a standard notion of fairness~\cite{foley1967resource, varian1974equity, stromquist1980cut} and the divide-and-choose method shows a fair division with respect to this notion can be found even if the second agent is secretive and just selects a piece \emph{after} we partition the cake. 

Asada et al.~\cite{asada2018fair} showed that this result extends to higher number of agents who are endowed with ordinal preferences: there exists a division of the cake into $n$ parts, which depends on the (subjective) preferences of only $n-1$ agents, such that the $n$th (secretive) agent can select an arbitrary piece and  still we would be able to allocate the remaining $n-1$ pieces in an envy-free manner. In other words, independent of the choice of the secretive agent, there exists an allocation in which no agent has a strictly stronger preference for the piece of any other agent (including the secretive one). This result of Asada et al.~\cite{asada2018fair} relies on a fixed-point argument--specifically, it uses a version of the Knaster, Kuratowski, and Mazurkiewicz (KKM) Lemma \cite{knaster1929beweis}. 

The work of Frick et al. \cite{frick2017achieving} shows that an envy-free \emph{rent division} can also be achieved, albeit inefficiently, in the absence of full information. Rent division is another well-studied problem in the fair-division literature, and it entails allocating $n$ rooms among $n$ agents and splitting the rent such that envy freeness is achieved; see, e.g., \cite{su1999rental}, \cite{svensson1983large}, \cite{alkan1991fair}, \cite{aragones1995derivation}, \cite{klijn2000algorithm} and \cite{Gal2017}. Considering agents who have ordinal preferences over the rooms (for every possible division of the rent), Frick et al. \cite{frick2017achieving} proved that the preferences of $n-1$ agents suffice to find a rent division such that secretive agent can select an arbitrary room (at its price) and still an overall envy-free allocation of the remaining $n-1$ rooms (at their respective prices and among the $n-1$ agents) would exist. 
 
In this paper we will consider agents who have numerical valuations over the good(s) (i.e., have cardinal preferences) and develop efficient algorithms for secretive fair division in multiple settings. Specifically, we will study fair division of rent, cake, and indivisible goods. We will show that, under appropriate valuations, for all of these settings a fair (or an approximately fair) division among $n$ agents can be efficiently computed using only the valuations of $n-1$ agents. The $n$th (secretive) agent can make an arbitrary selection after the division has been proposed and, irrespective of her choice, the computed division will continue to admit a fair allocation. 

Overall, this work establishes existence and algorithmic results for secretive analogues of standard fair-division problems.  \\

%the computed division will continue to admit a fair allocation among the remaining agents. 
%\begin{itemize}
\noindent
{\bf Fair Rent Division:} This problem entails splitting the rent and allocating $n$ rooms among $n$ agents such that, under the imposed rents, each agent prefers the room allocated to it over that of any other agent. In the cardinal version of this problem the preferences of the agents for the rooms are expressed via functions (one for each agent-room pair) which specify agents' utility for the rooms at every possible room rent. The works of Svensson \cite{svensson1983large} and Alkan et al. \cite{alkan1991fair} shows that if the utility functions of all the $n$ agents for each of the $n$ rooms are continuous, strictly decreasing, and bounded, then an envy-free rent division is guaranteed to exist. We extend these results and prove that, considering the utility functions of $n-1$ agents,\footnote{As in the standard rent-division case, these utilities are assumed to be continuous, strictly decreasing, and bounded.} we can propose a rent division among the $n$ rooms such that for every possible room selection of the secretive agent there exists an allocation (of the remaining $n-1$ rooms among the $n-1$ agents) which ensures overall envy freeness (Theorem \ref{theorem:existence}). Hence, analogous to the results of \cite{asada2018fair} and \cite{frick2017achieving} (which considered ordinal preferences), we show that fair rent division with a secretive agent can be achieved in the cardinal setting as well.\footnote{The assumptions considered in~\cite{asada2018fair} and~\cite{frick2017achieving} render these results incomparable with the cardinal setting of the present paper. In particular, these existence results require that every agent always prefers a zero-rent room over  rooms with positive rent. Since this assumption does not hold even for the class of quasilinear utilities, the ordinal setup of~\cite{asada2018fair} and~\cite{frick2017achieving} is incomparable with the cardinal-utility model considered in this work.}

Efficient algorithms for fair rent division are known for the quasilinear utility model \cite{aragones1995derivation, Gal2017, klijn2000algorithm}. In this setup every agent $a$ has a base value for each room $r$, and $a$'s utility for room $r$ when its rent is $p_r$ is equal to the base value minus $p_r$. We show that, under quasilinear utilities, an efficient, fair-division algorithm exists even in the secretive case, i.e., we develop a polynomial-time algorithm for secretive rent division when the utilities of the $n-1$ agents are quasilinear (Theorem \ref{theorem:secretive-ef}). \\

%The work of Su~\cite{} shows that such an envy-free rent division is guaranteed to exist when all the $n$ agents have ordinal preferences.
\noindent
{\bf Fair Division of Divisible Goods:} Cutting a cake (i.e., dividing a divisible good) in a fair manner is a classic problem in fair division. Formally, the cake is represented as the interval $[0,1]$ and it models resources that can be fractionally allocated, such as land. Each agent has a sigma additive valuation over the subintervals of $[0,1]$, i.e., over the pieces of the cake. The standard notions of fairness in this context are (i) envy freeness and (ii) proportionality. We develop results for both of these solution concepts: \\

\noindent
(i) \emph{Envy Freeness:} Simmons (see also \cite{su1999rental}) has shown that an envy-free partition of the cake among the $n$ agents is guaranteed to exist. However, no finite-time algorithm can find such a partition if the valuations are adaptively specified by an adversary \cite{stromquist2008envy}. Prior work has also considered a variant in which the cake is divided into more than $n$ pieces and disjoint subsets of the pieces are allocated among the agents. The best-known algorithm for finding an exact envy-free allocation with noncontiguous pieces has hyper-exponential complexity \cite{aziz2016discrete}. To address these computational barriers, one can consider a natural relaxation of envy freeness wherein for each agent $a$ the value of the piece (or bundle) assigned to it is within $\varepsilon$ of the value of the piece (or bundle) most preferred by $a$; here, we conform to the standard assumption that each agent's valuation for the entire cake is equal to one. Deng et al.~\cite{deng2012algorithmic} have shown that, for a small enough $\varepsilon$, finding an approximately envy-free contiguous $n$-partition of the cake is ${\rm PPAD}$-complete under Su's \cite{su1999rental} ordinal valuations model. 

Given these hardness results, we focus on the problem of dividing the cake into noncontiguous pieces and allocating bundles to agents such that approximate envy freeness is achieved. We show that in this framework a secretive division can be obtained efficiently. We consider this problem in the standard Robertson-Webb query model, and show that the query and time complexity of our algorithm is polynomial in the number of agents and $1/\varepsilon$, here $\varepsilon$ is the approximation parameter (Theorem~\ref{theorem:efcake}).  

In particular, considering the valuation of only $n-1$ agents, we can partition the cake into $n$ disjoint subsets (each potentially consisting of multiple pieces)  such that the secretive agent can select an arbitrary subset and we would still be able to allocate the remaining $n-1$ subsets in an approximately envy free manner. \\

\noindent
(ii) \emph{Proportionality:} In the context of cake cutting proportionality is another fundamental fairness criterion. It deems a partition to be fair if each agent receives a piece of value at least $1/n$ times her value for the entire cake \cite{steinhaus}. Note that additivity of the valuations ensure that an envy free allocation satisfies  proportionality. However, in contrast to envy freeness, a proportional division of the cake can be efficiently achieved by following the moving-knife procedure of Dubins and Spanier \cite{dubins1961cut}: a knife is moved from one end of the cake (i.e., the interval $[0,1]$) to the other and the cake is cut as soon as, for any agent $a$ (who does not have a piece yet), the value of the piece to the left of the knife is $1/n$ times the total value of the cake. This piece is allocated to agent $a$ and she is removed from consideration. The last agent gets the remainder of the cake. The fact that the agents' valuations are additive imply that the resulting allocation satisfies proportionality. We show that, interestingly, this moving-knife procedure can be executed with $n-1$ agents to obtain an $n$-partition which is proportionally fair even with a secretive agent (Theorem~\ref{theorem:propcake}).  \\

\noindent
{\bf Fair Division of Indivisible Goods:}  As mentioned previously, envy-free divisions always exist for divisible goods. By contrast, such a universal existential guarantee does not hold in the context of indivisible goods.\footnote{For example, consider a single indivisible good and two agents with nonzero valuation for the good. In any allocation, the losing agent will be envious.} To address this issue and motivated by allocation problems that involve discrete resources (such as courses at universities \cite{budish2016course}), a number of recent results have formulated and studied fairness notions that specifically address discrete goods \cite{budish2011combinatorial, procaccia2014fair, bouveret2016characterizing}. Two prominent notions in this line of work are (i) envy freeness up to one good ($\EFone$) and (ii) maximin shares ($\MMS$). We show that fairness in terms of these concepts can be achieved even in the presence of a secretive agent.  \\

\noindent
(i) $\EFone$ is a compelling analogue of envy freeness in the discrete setting \shortcite{budish2011combinatorial}. Specifically, an allocation is said to be $\EFone$ if every agent prefers her own bundle over any other agent's bundle, up to the removal of one good from the other agent's bundle. Lipton et al. \shortcite{lipton2004approximately} proved that if the valuations of the agents (over subsets of goods) are monotone, then an $\EFone$ allocation always exists and can be computed in polynomial time. The result is notably general, since it encompasses all monotone (combinatorial) valuations. We prove that, even with a secretive agent, fairness in terms of $\EFone$ can be guaranteed. In particular, we show that, given $n-1$ agents with monotone valuations, we can efficiently partition the set of indivisible goods into $n$ disjoint subsets, $\mathcal{S} = \{S_1, S_2, \ldots, S_n \}$, such that every collection of $n-1$ subsets from $\mathcal{S}$ can be assigned (among the $n-1$ agents) to obtain an $\EFone$ allocation overall (Theorem~\ref{theorem:ef1secretive}). Note that our result shows that an $\EFone$ allocation exists if $n-1$ agents have monotone valuations; the valuation of the $n$th agent can be completely arbitrary. Hence, as a corollary, we get a strengthening of the existence guarantee of Lipton et al. \shortcite{lipton2004approximately}. \\

\noindent %\footnote{In the case of indivisible goods, proportional fair divisions are not guaranteed to exist.}
(ii) $\MMS$: In the context of indivisible goods, maximin shares lead to another relevant notion of fairness. These shares provide an agent specific fairness threshold; in particular, the maximin share of an agent is defined to be the maximum value that the agent can ensure for herself if she were to (hypothetically) partition the set of indivisible goods into $n$ disjoint subsets and, then, from them receive the minimum valued one. An allocation is said to satisfy the maximin share guarantee ($\MMS$) if and only if each agent receives a bundle of value at least as much as her maximin share. Defined by Budish \cite{budish2011combinatorial}, maximin share can be interpreted through a discrete generalization of the cut-and-choose method: an agent $a$ partitions (cuts) the goods and the other agents get to select a bundle (choose) before $a$. A risk-averse agent $a$ would partition the goods to maximize the minimum valued bundle. Here, the value that $a$ can ensure for herself corresponds to her maximin share. 

Even though $\MMS$ allocations are not guaranteed to exist \cite{procaccia2014fair, kurokawa2016can}, this notion lends well to approximations: Procaccia and Wang \cite{procaccia2014fair} have shown that if the agents have additive valuations, then there exists an allocation wherein each agent receives a bundle of value at least $2/3$ times her maximin share; see also \cite{amanatidis2017approximation} and~\cite{barman2017approximation}. The work of Ghodsi et al. \cite{ghodsi2018fair} provides an efficient algorithm for finding a $3/4$-approximate $\MMS$ allocation under additive valuations. Approximation guarantees for $\MMS$ have also been obtained for more general valuation classes~\cite{barman2017approximation, ghodsi2018fair}. In particular, Ghodsi et al. \cite{ghodsi2018fair} show that a $1/3$-approximate $\MMS$ allocation can be computed efficiently under submodular valuations and a logarithmic approximation exists when the valuations are subadditive. 

We show that, even with a secretive agent, if the remaining $n-1$ agents have submodular valuations, then a $1/19$-approximate $\MMS$ allocation for them can be computed in polynomial time. In particular, we develop a combinatorial algorithm that finds a $1/19$-approximate $\MMS$ allocation by reducing the secretive problem to its standard fair-division version. Furthermore, we show that,  in the case of indivisible goods and additive valuations, a discrete extension of the moving-knife produce provides a $1/2$-approximate $\MMS$ allocation in the secretive case. \\

\noindent 
{\it Remark:} Note that, when considering envy freeness or $\EFone$ with a secretive agent, fairness is guaranteed for every agent. In particular, the secretive agent can always select its most preferred room/piece/bundle and achieve envy freeness, irrespective of its valuation type. This property holds in the $\MMS$ case as well, if the valuation of the secretive agent is additive: the secretive agent can select a bundle that ensures proportionality for her and, hence, obtain a bundle of value at least as much as her maximin share. However, in the $\MMS$ case such a guarantee cannot be achieved for a secretive agent whose valuation is, say, non-additive. This follows from the fact that, considering just the valuations of the $n-1$ non-secretive agents, one cannot determine a partition wherein some bundle is of value (with respect to the arbitrary valuation of the secretive agent) close to the maximin share of the secretive agent. \\

Overall, the paper shows that a broad spectrum of fair division problems can be addressed in a partial information setting. Indeed, with respect to a specific agent, these results prove that a limited number of valuation queries suffice to find a fair division. Furthermore, our work implies stronger versions of standard existential and algorithmic results, e.g., the rent-division result developed in the paper shows that as long as $n-1$ agents have quasilinear utilities, an envy-free rent division can be computed efficiently. Here, the utility function of the $n$th agent can be arbitrary. This is in contrast to prior works, which require all the agents to have quasilinear utilities. We obtain these results by developing connections between different fair division paradigms, e.g., we establish the result for $\EFone$ by using the rent-division guarantees. These connections are mentioned below and they might be of independent interest.  \\

\noindent 
{\bf Our Techniques:} We use the generalization of the KKM lemma developed in Asada et al. \shortcite{asada2018fair} to establish the existence of fair rent division with a secretive agent. % It is relevant to note that the work of Frick et al. \shortcite{frick2017achieving} (which addresses ordinal preferences) does not encompass the cardinal setting considered in this paper. In particular, the existence guarantee in Frick et al. \shortcite{frick2017achieving} requires that each agent prefers a room with zero rent over any room with nonzero rent. Since, in general, this assumption is not satisfied even for quasilinear utilities, the result of Frick et al. \shortcite{frick2017achieving} does not directly hold for cardinal preferences.  

Our efficient algorithm for quasilinear utilities draws upon the equivalence between envy-free rents and Walrasian equilibria. Specifically, the Second Welfare Theorem for Walrasian equilibrium ensures that if the $n-1$ agents have quasilinear utilities, then every maximum weight matching between the agents and the rooms induces an envy-free solution; here the weights between agents and rooms are set to be the corresponding base values. We use this property to  formulate a linear program whose solution is a fair rent division which is robust to the choice of the secretive agent.

Instantiating the result for quasilinear utilities, we show that if $n-1$ agents have monotone valuations, then an $\EFone$ allocation (with a secretive agent) always exists and can be computed efficiently. Specifically, we develop an algorithm that iteratively allocates goods to bundles and maintains the $\EFone$ property as an invariant. We use the rent-division result to show that in each iteration the algorithm can efficiently find a bundle and allocate the next good to it such that the invariant continues to hold. Note that, in contrast to the existential result of Lipton et al.---which relies on a direct parity argument---our work uses a novel connection between rent division and discrete fair division. 

The developed $\EFone$ algorithm in turn provides an efficient method to find an approximate envy-free cake division through a reduction of Lipton et al. \cite{lipton2004approximately}: we partition the cake into pieces such that, for each of the $n-1$ agents, the value of any piece is at most $\varepsilon$. Then, considering these low-valued pieces as indivisible goods, we execute the $\EFone$ algorithm to obtain an approximately envy-free division which can accommodate an arbitrary selection by the secretive agent.

We show that a careful analysis of a modified version of the moving knife procedure of Dubins and Spanier \shortcite{dubins1961cut} guarantees a proportional division of the cake for every eventual choice of the secretive agent.

To solve the problem of computing approximate $\mms$ allocations in the presence of a secretive agent, we employ a discrete moving knife procedure on top of existing algorithmic results (that were developed for the non-secretive setting).

\section{Notation and Preliminaries}
\label{section:problems}

\subsection{Rent Division}

{\bf Envy-Free Rent Division:} An instance of the (standard) envy-free rent division problem is represented by the following tuple $\langle \mathcal{A}, \mathcal{R}, \{ v_a(r, \cdot) \}_{a\in \mathcal{A}, r \in \mathcal{R}} \  \rangle$ wherein $\mathcal{A} :=[n]$  denotes the set of $n$ agents and $\mathcal{R} := [n]$ denotes the set of $n$ rooms. The cardinal preference of each agent $a \in \mathcal{A}$ for every room $r \in  \mathcal{R}$ is specified via a utility function $v_a(r, \cdot)$, i.e., $a$'s utility for $r$ at price (rent) $p_r \in \mathbb{R}$ is $v_a(r, p_r) \in \mathbb{R}$.

A solution to the rent division problem comprises of the tuple $(\pi,p)$, where $\pi: \mathcal{A} \mapsto \mathcal{R}$ is a bijection from the set of agents to the set of rooms and $p \in \mathbb{R}^n$ is a price vector for the set of rooms. The tuple $(\pi,p)$ is said to be an \emph{envy-free solution} of the given instance if for all $a \in \mathcal{A}$ and $r \in \mathcal{R}$ we have $v_a(\pi(a), p_{\pi(a)}) \geq v_a(r,p_r)$, i.e., under the given price vector, no agent strongly prefers (envies) any other room to the one allocated to her. \\

\noindent
{\bf Envy-Free Rent Division with a Secretive Agent:} A rent-division instance with a secretive agent is a tuple $\langle \mathcal{A}, \mathcal{R}, \{ v_a(r, \cdot) \}_{a\in \mathcal{A}\setminus \{n\}, r \in \mathcal{R}} \rangle$, wherein $\mathcal{A} =[n]$ and $\mathcal{R}=[n]$ denote the set of $n$ agents and  $n$ rooms, respectively. We will use $v_a(r, \cdot)$ to denote the utility function of agent $a \in \mathcal{A} \setminus\{n\}$ for room $r \in \mathcal{R}$. Here, in contrast to the standard rent division setting, the problem instances have a distinguished, ``secretive'' agent, $n$, who gets to pick any room of her choice after the prices for the rooms have been proposed.\footnote{Note that, here, the instances do not contain any information about the utilities of the ($n$th) secretive agent.} A secretive envy-free solution is a  price vector $p \in \mathbb{R}^n$ that satisfies the following property: for every room $k \in \mathcal{R}$, there exists a bijection $\pi_k: \mathcal{A} \setminus \{n\} \mapsto \mathcal{R}\setminus{\{k\}}$ such that $(\pi_k, p)$ is envy-free, i.e., for all $a \in \mathcal{A} \setminus \{n\}$ and $r \in \mathcal{R}$ we have $v_a(\pi_k(a),p_{\pi_k(a)}) \geq v_a(r, p_r)$.

%of the instance $\langle \mathcal{A}, \mathcal{R}, \{ v_a(r, \cdot) \}_{a\in \mathcal{A}\setminus\{n\}, r \in \mathcal{R}} \rangle$ 

%%%
Intuitively, a price vector $p \in \mathbb{R}^n$ constitutes a solution in the secretive setting if for any arbitrary room choice, $k \in \mathcal{R}$, of the secretive agent, the properties of $p$ allow us to allocate the remaining rooms (using $\pi_k$) among the agents such that no agent strictly prefers any other agent's room. Note that such a price vector $p$ is computed \emph{before} the choice $k$ is made by the secretive agent and without taking into account any information about her utilities.

We will establish existential results for bounded, continuous, and monotone (strictly)  decreasing utilities. The utility functions $\{ v_a(r, \cdot) \}_{a\in \mathcal{A}\setminus\{n\}, r \in \mathcal{R}}$ are bounded in the sense that there exists $M \in \mathbb{R_+}$ such that for all agents $a \in \mathcal{A} \setminus \{n\}$ and all pairs of rooms $r, r' \in \mathcal{R}$ the following inequality holds. 
\begin{align*}
v_a(r,M) < v_a(r',0)
\end{align*}

Quasilinear utilities constitute a well-studied subclass of the above-mentioned (broad) class of utility functions. In the quasilinear setting each utility function $v_a(r,x)$ is of the form $B^a_r - x$; here $B^a_r \in \mathbb{R}_+$ is the base value of room $r$ for agent $a$. In other words, in the quasilinear case the utility of agent $a$ for room $r$ is equal to her base value for the room minus the rent of $r$.

\subsection{Envy-Free Allocation of Goods}

%,  i.e., for any subset $S \subset \mathcal{G}$ and good $g \in \mathcal{G}$ we have  $v_a(S \cup \{g\}) \ge v_a(S)$
\noindent
{\bf Secretive $\EFone$ Allocations of Indivisible Goods:} An instance of the fair division problem with indivisible goods and a secretive agent is a tuple  $\langle \mathcal{A}, \mathcal{G}, \{ v_a \}_{a\in \mathcal{A}\setminus \{n\}}  \rangle$ in which  $\mathcal{A} :=[n]$  stands for the set of $n$ agents, $\mathcal{G} :=[m]$ denotes the set of $m$ indivisible goods, and $v_a$s specify the valuation of each agent $a \in \mathcal{A}\setminus \{n\}$ over the set of goods. The valuation functions $v_a: 2^{\mathcal{G}} \mapsto \mathbb{R_+}$ are assumed to be nonnegative and monotone for all agents $a \in \mathcal{A}\setminus \{n\}$.\footnote{Analogous to the rent-division case, the instance tuple contains no information about the valuation function of agent $n$, who is designated to be the secretive agent.} Write $\Pi_n(\mathcal{G})$ to denote the set of all $n$-partitions of $\mathcal{G}$.

In the standard (non-secretive) context with $n$ agents and set of indivisible goods $\mathcal{G}$, an $n$-partition $\mathcal{A}=(A_1, \ldots, A_n) \in \Pi_n(\mathcal{G})$ is said to be \emph{envy-free up to one good} ($\EFone$) iff for all agents $a, b$ there exists a good $g \in A_b$ such that $v_a(A_a) \geq v_a(A_b \setminus \{ g \})$; here, each agent $a$ is assigned the subset of goods (bundle) $A_a$. Note that the definition ensures that each agent prefers its own bundle over the bundle of any other agent up to the removal of one good. 

This definition extends to address fair division of indivisible goods with a secretive agent. Formally, an $n$-partition $\mathcal{P} = (P_1,...,P_n) \in \Pi_n(\mathcal{G})$ of the set of goods $\mathcal{G}$ is said to be a secretive $\EFone$ allocation iff it satisfies the following property: for each choice of bundle $P_k$ from partition $\mathcal{P}$, there exists a bijection $\pi_k: \mathcal{A} \setminus \{n\} \mapsto [n]\setminus{\{k\}}$ such that the allocation defined by $\pi_k$ on $(P_1, \ldots, P_{k-1}, P_{k+1}, \ldots, P_n)$ is $\EFone$, i.e., for all $a \in \mathcal{A} \setminus \{n\}$ and all bundles $P_b$, with $b \in [n]$, there exists a good $g \in P_b$ such that $v_a(P_{\pi_k(a)}) \geq v_a(P_b \setminus \{g\})$.
%\begin{align*}
%\end{align*} 

%%%
In other words, for every choice of bundle $P_k$ made by the secretive agent, the secretive $\EFone$ solution $\mathcal{P} = (P_1,P_2,...,P_n)$ allows for an $\EFone$ allocation of the remaining bundles among the first $n-1$ agents. \\

%Our contribution is a polynomial time algoroithm to find a secretive $\EFone$ solution of a given instance. 

%Our contribution is a polynomial time algoroithm to find a secretive $\EFone$ solution of a given instance. \\
\noindent
{\bf Secretive Envy-Free Cake Cutting:} A cake-cutting instance with a secretive agent comprises of a cake, represented by the unit interval $ C =[0,1]$, the set of agents, represented by $\mathcal{A} :=[n]$, and valuations $\{v_a\}_{a \in \mathcal{A}\setminus \{n\}}$. These valuation functions are defined over all finite unions of disjoint intervals of $C$. Following standard conventions, the functions are assumed to be nonnegative, normalized (i.e., $v_a(C) =1$) and sigma additive.\footnote{That is, we have $v_a(X\cup Y ) = v_a(X) + v_a(Y)$ for all disjoint pairs of  intervals $X,Y \subseteq C$.} As is typical in cake cutting, the valuations are also assumed to be divisible, i.e., for all intervals $X \subseteq C$ and  real numbers $\lambda \in (0,1]$, there exists a subinterval $X' \subseteq X$ such that $v_a(X') = \lambda v_a(X)$. Here, we use term bundle to denote a finite union of disjoint intervals of the cake. An $n$-partition of the cake $\mathcal{A} = (A_1,\ldots, A_n)$ is defined to be a tuple of $n$ bundles which are pairwise disjoint and whose union is equal to $C$. %We further refine this notion and will say that an $n$-partition of the cake, $(A_1,\ldots, A_n)$, is contiguous iff each $A_i$ is a single (contiguous) interval of $C$. 

Analogous to the previous settings, the valuations are specified only for the first $n-1$ agents and the $n$th agent is considered to be secretive--no information about her valuation function over the cake is specified.

%As mentioned previously, finding (exact) envy-free partitions is a challenging problem. Hence, this work considers a natural approximation.
In this work we consider the problem of finding an approximately envy-free partition of the cake. In particular, we say that an allocation is $\varepsilon$-envy free ($\epsEF$) if, under it the envy is upper bounded by $\varepsilon$, i.e., an $n$-partition $\mathcal{A} = (A_1,\ldots,A_n)$ is $\epsEF$ iff, for all agents $a$ and indices $b \in [n]$, we have $v_a(A_a) \geq v_a(A_b) - \varepsilon$; here, for all agents $a$, bundle $A_a$ is allocated to $a$. 
This paper considers cake cutting in the standard the Robertson-Webb query model \cite{robertson1998cake}, which provides access to the following (query) oracles:

%This paper works the Roberston-Webb query model \cite{robertson1998cake}, 
\begin{enumerate}
	\item Cut($a,I,\lambda$): Given agent $a$, an interval $I = [x_1,x_2] \subseteq C$ of the cake, and a real number $\lambda \in (0,1]$ as a query, the response is a point $x \in I$ such that $v_a([x_1,x]) = \lambda v_a(I)$.  
	\item Eval($a,I$):  When queried with an agent $a$ and an interval $I = [x_1,x_2] \subseteq C$, this oracle provides $a$'s valuation of interval $I$, i.e., provides $v_a(I)$.
\end{enumerate}

Extending this concept, a partition $\mathcal{A} = (A_1,\ldots,A_n)$ of the cake $C$ is defined to be a \emph{secretive $\epsEF$} solution iff for every bundle $A_k$ in the partition $\mathcal{A}$, there exists a bijection $\pi_k: \mathcal{A} \setminus \{n\} \mapsto [n] \setminus {\{k\}}$ such that the allocation defined by $\pi_k$ on $(A_1,A_2,\ldots, A_{k-1}, A_{k+1}, \ldots, A_n)$ is $\epsEF$, i.e., for all agents $a  \in \mathcal{A} \setminus \{n\}$ and all bundles $A_b$ for $b\in [n]$ we have $v_a(A_{\pi_k(a)}) \geq v_a(A_b) - \varepsilon$. \\ 
%\begin{align*}
%v_a(A_{\pi_k(a)}) \geq v_a(A_j) - \varepsilon
%\end{align*} 

%\begin{remark2}
Note that the secretive agent gets to pick her first choice. Therefore, in all the settings mentioned above (cake cutting, division of indivisible goods, and rent division), the secretive agent can achieve envy freeness. 
%\end{remark2}

\subsection{Proportional Division of Goods}
{\bf Proportional Cake Division with a Secretive Agent:} We also obtain results for cake-cutting with a secretive agent, in terms of proportionality. Formally, an $n$-partition $\mathcal{A} = (A_1,A_2,...,A_n)$ of the cake $C$ is said to be a secretive proportional solution iff for every bundle $A_k$ of $\mathcal{A}$, there exists a bijection $\pi_k: \mathcal{A} \setminus \{n\} \mapsto [n]\setminus{\{k\}}$ such that the allocation defined by $\pi_k$ on $(A_1,A_2,\ldots, A_{k-1}, A_{k+1}, \ldots, A_n)$ is proportional, i.e., for all agents $a \in \mathcal{A} \setminus \{n\}$ we have $v_a(A_{\pi_k(a)}) \ge 1/n$.\footnote{Recall that the valuations of the first $n-1$ agents are normalized to satisfy $v_a(C) = 1$.} For proportional fairness, we in fact show that there exists a contiguous $n$-partition that achieves proportional fairness even with a secretive agent. \\

\noindent
{\bf $\MMS$ Allocation of Indivisible Goods with a Secretive Agent:} As mentioned in Section~\ref{section:introduction}, the maximin share property ($\MMS$) provides an analogue of proportionality in the context of indivisible goods. Given a set of indivisible goods $\mathcal{G}$ and $n$ agents, the maximin share, $\mu_a$, of an agent $a$ is defined to be the maximum value that $a$ can guarantee for herself, if she were to partition $\mathcal{G}$ into $n$ subsets and then receive the minimum valued one. Formally, with $v_a: 2^\mathcal{G} \mapsto \mathbb{R}_+$ representing the valuation of agent $a$ over the set of indivisible goods $\mathcal{G}$, we have  
\begin{align}
\label{eq:mms-defn}
	\mu_a := \max_{(A_1,\dots,A_n) \in {\Pi_n(\mathcal{G})}} \  \min_{j \in [n]} v_a(A_j)
\end{align}

Since $\MMS$ allocations---i.e., allocations in which each agent gets a bundle of value at least as large as her maximin share---do not always exist, prior work has focused on approximation guarantees. The goal here is to find a partition $(A_1,A_2,...,A_n)$, wherein each agent $a$ is allocated bundle, $A_a$, of value (under $v_a$) at least $\alpha \mu_a$, with parameter $\alpha \in (0,1]$ being as large as possible. 

In the presence of a secretive agent, the problem instance for maximin fairness is a tuple $\langle \mathcal{A}, \mathcal{G}, \{ v_a \}_{a\in \mathcal{A}\setminus \{n\}}  \rangle$; here $\mathcal{A} =[n]$ denotes the set of agents, $\mathcal{G} = [m]$ denotes the set of indivisible goods, and the valuation of each (non-secretive) agent $a \in \mathcal{A} \setminus \{n\}$ is represented by $ v_a : 2^\mathcal{G} \mapsto \mathbb{R}_+$.

An $n$-partition $\mathcal{P} = (P_1,P_2,...,P_n) \in \Pi_n(\mathcal{G})$ is defined to be a \emph{secretive $\alpha$-$\MMS$} solution iff for each bundle $P_k$ of partition $\mathcal{P}$, there exists a bijection $\pi_k: \mathcal{A} \setminus \{n\} \mapsto [n]\setminus{\{k\}}$ such that the allocation defined by $\pi_k$ on $(P_1, P_2, \ldots, P_{k-1}, P_{k+1}, \ldots, P_n)$ gives each agent $a \in \mathcal{A} \setminus \{n\} $ a bundle of value $\alpha$ times her maximin share, $\mu_a$ , i.e., for all $a \in \mathcal{A} \setminus \{n\}$ the following inequality holds $v_a(A_{\pi_k(a)}) \geq \alpha \mu_a$ .

For (approximate) maximin fairness, the present work focuses on valuations that are monotone, nonnegative, and submodular. Recall that a function $f: 2^{\mathcal{G}} \mapsto \mathbb{R}_+$  is said to be submodular iff it satisfies the diminishing marginals property, i.e., for all subsets  $A, B \subseteq \mathcal{G}$, with $A \subseteq B$, and every $g \in \mathcal{G} \setminus B$ we have  $f(A \cup \{g\}) -f(A) \geq f(B \cup \{g\}) -f(B)$.

%We are specifically interested in submodular valuations, which we will focus on in this paper. A valuation function  %We will denote the marginal value of good $g$ with respect to set $A$ (i.e $f(A \cup \{g\}) -f(A)$ ) by $f_a

\subsection{Framework for Secretive Solution Concepts}

In all the solution concepts defined above, the bijections/permutations $\{\pi_1,\ldots,\pi_n\}$ (that enable a fair division after the secretive agent makes a choice) are not an explicit part of the solution. However, for all settings considered in this work, these permutations can be computed efficiently.  %Throughout the rest of the paper, we will refer to these $\pi_i$s as the permutations associated with a secretive solution and will  outline the computation of these permutations while establishing our algorithmic results.

%Add Storng Hall's Interpreatation

%We will use $\mathbb{G}$ to denote the complete bipartite graph between the agents and the bundles, $\mathbb{G} := (\mathcal{A} \cup \mathcal{P}, \mathcal{A} \times \mathcal{P})$. Let 

We now provide a useful formalism for secretive solutions. % which, in particular, will be used for finding secretive $\EFone$ allocations.  
Let $\mathcal{P} = (P_1,\ldots,P_n)$ represent a collection of $n$ bundles that constitute a candidate solution of a fair division problem with agents $\mathcal{A}$ and known valuations $\{v_a\}_{a \in \mathcal{A} \setminus \{n\}}$. In this abstract setting, the bundles can be divisible or indivisible goods, or even a combination of both (e.g., a bundle can represent a room and its rent). 

Write $\mathcal{H}$ to denote the bipartite graph with the two vertex parts being $\mathcal{A} \setminus \{n \}$ and  $\{P_1, P_2, \ldots, P_n\}$, respectively.\footnote{Note that the two bipartite vertex sets of $\mathcal{H}$ are of size $n-1$ and $n$, respectively.} The edge set of $\mathcal{H}$ is constructed to account for the underlying fairness property: edge $(a,P_i)$ is included in $\mathcal{H}$ iff bundle $P_i$ satisfies the fairness criterion for agent $a$. For example, in the context of rent division---wherein each bundle $P_i$ is a tuple of a room $i$ and its rent $p_i$---edge $(a, P_i)$ is included iff assigning room $i$ to agent $a$ ensures envy freeness for $a$, i.e., iff $v_a(i, p_i) \geq v_a(r, p_r)$ for $r \in \mathcal{R}$. 

Note that the collection of bundles $\mathcal{P}$ is a secretive solution (satisfying the required fairness criterion) iff for every $k \in [n]$ (i.e., each choice $P_k$ of the secretive agent), there exists a perfect matching $\pi_k$ in the bipartite graph obtained by removing (vertex) $P_k$ from $\mathcal{H}$. This property leads to the following characterization: a collection of bundles $\mathcal{P}$ is a secretive solution, with respect to the underlying fairness property, iff for every subset $S \subseteq \mathcal{A} \setminus \{n\}$ we have $|\Gamma_\mathcal{H}(S)|  \geq |S| +1$; here $\Gamma_\mathcal{H}(S)$ denotes the set of neighbors of $S$ in the graph $\mathcal{H}$. %We will refer to this property as the strong Hall's condition, since it is a strengthening of the condition in Hall's theorem.

%which provides a necessary and sufficient condition for the existence of perfect matchings in bipartite graphs. 

%That is, we obtain the vertex set of $\mathbb{G}_k$ by removing the $n$th agent from $\mathcal{A}$ and the $k$th bundle, $P_k$ from $\mathcal{P}$. 
%Furthermore, with respect to a fairness property $\mathbb{P}$ (e.g., envy freeness), write $\mathcal{X}_\mathP$ represent the set of edges $(a,P_i)$ that satisfy  Let $\mathcal{X}_P$ represent the set of edges $(a,P_i)$ that satisfy a desirable property $\mathcal{Q}$. We will use the above characterization in later sections to . 

The following lemma establishes another useful result regarding the construction of the $n$ permutations $\{\pi_1,\ldots,\pi_n\}$. % from a particular type of mapping. 
In this result the mapping $\sigma$ can be thought of as a ``backup scheme:'' to ensure that a tuple $(P_1, \ldots, P_n)$ is a secretive solution is suffices to show that, for each agent $a \in \mathcal{A} \setminus \{n\}$, both the bundles $P_a$ and $P_{\sigma(a)}$, with $\sigma(a) > a$, satisfy the fairness property. The fact that $\sigma(a)> a$, for all the first $n-1$ agents, implies that $\sigma$ induces a topological ordering, i.e., it is acyclic. Intuitively, with mapping $\sigma$ in hand, we can either given an agent $a$ the bundle $P_a$ or, in case $P_a$ is not available, we can allocate $P_{\sigma(a)}$ to $a$. The acyclicity of $\sigma$ ensures that each agent gets a different bundle. 

\begin{lemma}
	\label{lemma:backup}
Given a mapping $\sigma : [n-1] \mapsto [n]\setminus \{ 1 \}$ with the property that $\sigma(a) > a$, for all $a \in [n-1]$, we can efficiently find a set of $n$ bijections $\mathcal{B} =  \{\pi_k: [n-1] \mapsto [n]\setminus{\{k\}} \}_{k \in [n]}$ such that for all the bijections $\pi \in \mathcal{B}$ and every $a \in [n-1]$ we have $\pi(a) \in \{a, \sigma(a)\}$, i.e., for each $a$, permutation $\pi$ is either the identity or $\sigma$.    
\end{lemma}

\begin{proof}
Consider the directed graph $\mathcal{J}:= \left( [n], \{ (i,\sigma(i))\}_{i \in [n-1]} \right)$. In $\mathcal{J}$, the $n$th vertex has no outgoing edge and the outdegree of every other vertex is exactly equal to one. In addition, the fact that $\sigma(a) > a$, for all $a \in [n-1]$, implies that $\mathcal{J}$ is acyclic. Therefore, for each $k \in [n-1]$, there exists a directed path, $S_k$, from vertex $k$ to the last vertex $n$. The required bijection $\pi_k$ is constructed in the following manner: for all vertices $i$ not on the path $S_k$ we set $\pi_k(i) = i$ and for the remaining vertices set $\pi_k(i) =\sigma(i)$. 
\end{proof}

%The above result has an important interpretation in context of secretive fair division. The mapping $\sigma$ can be thought of as a backup allocation scheme. The above lemma states that it is sufficient to show the existence of a cycle-free backup scheme to prove the existence of the $n$ associated bijections of a secretive solution.

\section{Main Results}
The statements of our key results are provided in this section. \\

\noindent
{\bf Fair Rent Division:} The existential and algorithmic results for fair rent division with a secretive agent are established in Section \ref{section:rent}.

\begin{restatable}{theorem}{TheoremExistence}
\label{theorem:existence}
Let $\mathcal{I} = \left\langle \mathcal{A}, \mathcal{R},  \{ v_a(r, \cdot) \}_{a\in \mathcal{A}\setminus \{n\}, r \in \mathcal{R}} \right\rangle$ be a rent-division instance with a secretive agent. If the utilities, $\{ v_a(r, \cdot) \}_{a\in \mathcal{A}\setminus \{n\}, r \in \mathcal{R}}$, of the instance are bounded, continuous, and monotone decreasing, then $\mathcal{I}$ admits a secretive envy-free solution $p \in \mathbb{R}^{n}$. 
\end{restatable}

\begin{restatable}{theorem}{TheoremRent}
\label{theorem:secretive-ef}
Given any rent-division instance $\mathcal{I} =\langle \mathcal{A}, \mathcal{R}, \{ v_a(r, \cdot) \}_{a\in \mathcal{A}\setminus \{n\}, r \in \mathcal{R}} \  \rangle$ with a secretive agent and quasilinear utilities, a secretive envy-free solution $p \in \mathbb{R}^{n}_{+}$ of $\mathcal{I}$ can be computed in polynomial time.
\end{restatable}

\noindent
{\bf Envy Free and $\EFone$ Allocations:} In Section \ref{section:envy} we address fairness in terms of envy and prove the following two results.

\begin{restatable}{theorem}{TheoremEFone}
\label{theorem:ef1secretive}
Let $\mathcal{I} = \langle \mathcal{A}, \mathcal{G}, \{ v_a \}_{a\in \mathcal{A}\setminus \{n\}} \  \rangle$ be a fair division instance with indivisible goods, $\mathcal{G}$, and a secretive agent. If the valuations $\{ v_a \}_{a \in \mathcal{A}\setminus \{n\}}$ are nonnegative and monotone (increasing), then a secretive $\EFone$ solution $\mathcal{P} = (P_1,P_2,\ldots,P_n)$ for $\mathcal{I}$ is guaranteed to exist and can be computed in polynomial time.  
\end{restatable}
The computational part of the previous result only requires oracle access to the underlying valuations (set functions) $\{ v_a \}_{a \in \mathcal{A}\setminus \{n\}}$.

\begin{restatable}{theorem}{TheoremEFCake}
\label{theorem:efcake}
For any cake-cutting instance $\mathcal{I} = \langle \mathcal{A}, C =[0,1], \{ v_a \}_{a\in \mathcal{A}\setminus \{n\}} \  \rangle$ with a secretive agent, a secretive $\epsEF$ solution $\mathcal{P} = (P_1, \ldots, P_n)$ can be computed in time that is polynomial in $n$ and $1/\varepsilon$, under the Robertson-Webb query model. 
\end{restatable}

\noindent 
{\bf Proportional and $\MMS$ Allocations:} Results for proportionality and $\MMS$ are developed in Section \ref{section:propmms}. 

\begin{restatable}{theorem}{TheoremPropcake}
\label{theorem:propcake}
For any cake cutting instance $\mathcal{I} = \langle \mathcal{A}, C =[0,1], \{ v_a \}_{a\in \mathcal{A}\setminus \{n\}} \  \rangle$ with a secretive agent, a secretive proportional solution $\mathcal{P} = (P_1,P_2,\ldots,P_n)$ can be computed in polynomial time under the Robertson-Webb query model. Moreover, we can ensure that the computed bundles, $P_i$s, are intervals, i.e., we can efficiently find a solution $\mathcal{P}$ which forms a contiguous partition of the cake. 
\end{restatable}

\begin{restatable}{theorem}{TheoremMMSsecretive}
	\label{theorem:mmssecretive}
Let $\mathcal{I} = \langle \mathcal{A}, \mathcal{G}, \{ v_a \}_{a\in \mathcal{A}\setminus \{n\}} \  \rangle$ be a fair division instance with indivisible goods, $\mathcal{G}$, and a secretive agent. If the valuations $\{ v_a \}_{a \in \mathcal{A}\setminus \{n\}}$ are nonnegative, monotone, and submodular, then  a secretive $1/19$-$\mms$ allocation $\mathcal{P} = (P_1,P_2,...,P_n)$ for $\mathcal{I}$ is guaranteed to exist and such an approximate $\MMS$ allocation can be computed in polynomial time.
\end{restatable}

\section{Envy-Free Rent Division}
\label{section:rent}
%This section presents existential and algorithmic results for envy free division of goods with a secretive agent in different settings. We prove the existence of secretive envy free solutions for rent division instances with reasonably general utility functions. Interestingly, this existence result directly leads to an algorithm that efficiently computes these solutions for rent division instances with quasilinear utilities.

This section presents our results for fair rent division. We show that, for any rent-division instance with reasonably general utility functions, a secretive envy-free solution is guaranteed to exist. We complement this existential result by developing an efficient algorithm for computing such solutions for when the known utilities are quasilinear.

\subsection{Existential Result}
%Intro- Existence then algorithm
%Cite Asada+
%Def KKM Cover
%Define h function
%Define the agent sets and prove they are KKM Covers
%Theorem and proof

This subsection considers rent-division instances $\mathcal{I} =\langle \mathcal{A} =[n], \mathcal{R}=[n], \{ v_a(r, \cdot) \}_{a\in \mathcal{A}\setminus \{n\}, r \in \mathcal{R}} \  \rangle$ with a secretive agent and utilities, $\{ v_a(r, \cdot) \}_{a\in \mathcal{A}\setminus \{n\}, r \in \mathcal{R}}$, that are bounded, continuous, and monotone decreasing. Employing a generalization of the Knaster-Kuratowski-Mazurkiewicz (KKM) lemma, one can prove that such rent-division instances always admit a secretive envy-free solution. This form of KKM lemma appears in the work of Asada et al. \shortcite{asada2018fair}, wherein it is used to prove the existence of secretive envy-free solutions under ordinal utilities. As mentioned previously, the result in Asada et al. \shortcite{asada2018fair} requires assumptions which renders it incomparable with the current (cardinal) setup; in particular, quasilinear utilities do not fall under the valuation classes considered in Asada et al. \shortcite{asada2018fair}.

Recall that for bounded utility functions $\left\{ v_a(r, \cdot)\right\}_{a \in \mathcal{A} \setminus \{n\}, r \in \mathcal{R}}$ there exist parameter $M \in \mathbb{R_+}$ such that for all agents $a \in \mathcal{A} \setminus \{n\}$ and all pairs of rooms $r,r' \in \mathcal{R}$ we have:
\begin{align}
\label{inequality:boundedness}
v_a(r,M) < v_a(r',0)
\end{align}
Write $\{e_i\}_{i \in [n]}$ to denote the standard basis vectors of $\mathbb{R}^n$ and let $\Delta^n$ be the standard simplex in $\mathbb{R}^n$, i.e., $\Delta^n:=\text{conv}( \{ e_i\}_{i\in [n]})$.

First we define KKM covers of the simplex and then state the KKM generalization of Asada et al. \shortcite{asada2018fair}. 
%We define KKM Covers of the standard $(n-1)$ dimensional simplex and state the result of Asada et al. ~\cite{}.
\begin{definition}[KKM Cover]
	\label{definition:kkmcover}
	A collection of $n$ closed sets $ C_1,C_2,...,C_n \subset \mathbb{R}^n $ is said to form a KKM cover of the simplex $\Delta^n$, if and only if for all $I \subseteq [n]$, the convex hull of the basis vectors corresponding to $I$ is covered by $\bigcup_{i \in I} C_i$, i.e., iff for all $I \subseteq [n]$, we have $\text{conv}\left(\{e_i\}_{i \in I}\right) \subseteq  \bigcup_{i \in I} C_i$.
\end{definition}
%Armed with the definition of a KKM cover, we can now state the lemma proved by Asada et al. ~\cite{}.
	The following lemma, proved in~\cite{asada2018fair}, provides a useful property of any $(n-1)$ KKM covers of $\Delta^n$. The lemma states that for every index $k \in [n]$, it is possible to pick one set from each of the $(n-1)$ KKM covers such that no two sets have the same index and the index $k$ is not used, with the property that the intersection of these selected sets is non empty.
	%\textcolor{red}{Add a short description of the lemma}
	\begin{lemma}[\cite{asada2018fair}]
		\label{lemma:strongcolourful}
		Given $(n-1)$ KKM Covers $\mathcal{C}^1, \mathcal{C}^2, \ldots, \mathcal{C}^{n-1}$ (here each $\mathcal{C}^i=\{C^i_1, C^i_2, \ldots, C^i_n \} $ is a collection of $n$ sets) of the standard simplex $\Delta^n$, there exists a point $x \in \Delta^n$ and $n$ bijections $\pi_k: [n-1] \mapsto [n] \setminus \{k\}$ with $1 \leq k \leq [n]$ such that $x \in \bigcap_{i \in [n-1]} C^i_{\pi_k(i)}$. 
	\end{lemma}

To apply this lemma we will map points in the simplex to room rents (price vectors).  Such a mapping enables us to construct KKM covers from the utilities of the first $n-1$ agents. Specifically, we define a function $h(z) := M(1- nz)$, where $M \in \mathbb{R}_+$ is a large enough real number that satisfies the boundedness condition (\ref{inequality:boundedness}) for the given rent-division instance. Given a simplex point $x \in \Delta^n$, we apply $h$ component-wise to generate the price (rent) of each room, i.e., price  vector $p$ is generated from $x$ by setting $p_r := h(x_r)$ for all $r \in [n]$. Note that $h(0)=M$ and $h(z) \leq 0$ for all $z \geq 1/n$.

For each agent $a \in \mathcal{A} \setminus \{ n \}$, we define a collection of sets $\mathcal{C}^a = \{ C^a_1,C^a_2,...,C^a_n \}$ such that the set $C^a_r$ consists of all the simplex points whose corresponding prices render room $r$ as a ``first choice'' of agent $a$. Formally,  
	\begin{align}
	\label{definition:kkmcovers}
	C^a_r := \{&x \in \Delta^{n} \mid v_a(r,h(x_r)) \ge v_a(s , h(x_{s})) \ \text{ for all } s \in \mathcal{R} \}.	
	\end{align}
	
	The following claim establishes that each agent's collection of sets forms a KKM cover. 
\begin{claim}
For each agent $a \in \mathcal{A} \setminus \{ n \}$, the set $\mathcal{C}^a = \{C^a_1, \ldots, C^a_n\}$ (as defined above) is a KKM cover of the simplex $\Delta^n$.
\end{claim} 
\begin{proof}
Since the valuation functions are continuous, using the sequential criteria for continuity, we get that the sets in the collection $\mathcal{C}^a$ are closed.

Now, consider any subset $I \subseteq [n]$ and any point $x \in \text{conv}(\{e_i\}_{i \in I})$, where  $e_i$s denote the standard basis vectors of $\mathbb{R}^n$ (these vectors are also the vertices of the simplex $\Delta^{n}$). To prove that $\mathcal{C}^a$ is a KKM cover, it suffices to prove that there exists $r \in I$ such that $x$ is contained in the set $C^a_r$. 

Write $p \in \mathbb{R}^n_+$ to denote the price vector obtained by (component-wise) applying function $h$ on $x$. Note that, for all $r' \in [n] \setminus I$, we have $x_{r'} = 0$. Hence, the definition of $h$ gives us $p_{r'} = M$ for all $r' \in [n] \setminus I$. On the other hand, an averaging argument implies that there exists a component, $\rho \in I$, of $x$ such that $x_\rho \geq 1/n$ and, hence, $p_\rho \leq 0$. Therefore, using the definition of $M$ (see inequality  (\ref{inequality:boundedness})) and the monotonicity of $v_a(r,\cdot)$s, we get that $v_a(r', p_{r'}) < v_a( \rho, p_\rho)$ for all $r' \in [n] \setminus I$. In other words, at price vector $p$, none of the rooms $r' \in [n] \setminus I$ is a first choice of agent $a$, $\left([n] \setminus I \right) \cap \left( \argmax_{r} v_a(r, p_r) \right)= \emptyset$. Hence, at price vector $p$, a first-choice room, say $r$, must be from the set $I$. Overall, this ensures that $x \in C^a_r$, for some $r \in I$, and the claim follows. 

% We know that some non empty set of rooms $F_a(p)$ constitutes the first choice set of agent $a$ at this price vector. By definition of the sets $\{C^a_r\}_{r \in [n]}$, we know that for all rooms $r \in F_a(p)$, the set $C^a_r$ contains the simplex point $x$. Thus, it suffices to prove that $F_a(p) \cap (I \setminus [n]) = \phi$. As $\text{Supp}(x) = I$, we know that for any $r' \in I \setminus [n]$, the component $x_{r'}$ is $0$. Further, as $x$ is in the unit simplex, there exists $r \in I$ such that $x_r \ge 1/n$. Comparing the utilities at room $r$ and any room $r' \in I \setminus [n]$ :
%	\begin{align*}
%		v_a(r,h(x_r)) &\ge v_a(r,0) \qquad &\text{(Montonicity of $h$ and $v_a(r,\cdot)$)} \\
%		&> v_a(r',M) \qquad &\text{(Definition of Boundedness - Inequality \ref{inequality:boundedness})} \\
%		&= v_a(r',h(x_{r'})) \qquad &\text{(Definition of function $h$)}
%	\end{align*}	
%	Thus, no room $r' \in I \setminus [n]$ can be a first choice of agent $a$ at point $x$, proving that $C^a$ is a KKM Cover.
\end{proof}

	With the $(n-1)$ KKM covers (one for each agent $a \in \mathcal{A} \setminus \{ n \}$) in hand, we apply Lemma \ref{lemma:strongcolourful} to find a secretive envy-free solution. 
 
\TheoremExistence*
\begin{proof}
Consider the $(n-1)$ KKM covers, $\mathcal{C}^1, \ldots, \mathcal{C}^{n-1}$ (obtained via Equation (\ref{definition:kkmcovers})) corresponding to the  valuation functions $\{ v_a(r, \cdot) \}_{a\in \mathcal{A}\setminus \{n\}, r \in \mathcal{R}}$. Lemma \ref{lemma:strongcolourful} ensures that there exists a point $x^* \in \Delta^{n}$ and $n$ bijections $\pi_k: [n-1] \mapsto [n] \setminus \{k\}$, with $1 \leq k \leq n$, such that $x^* \in \bigcap_{i \in [n-1]} C^i_{\pi_k(i)}$. We will show that the   price vector, $p^*$, obtained by componentwise applying function $h$ on $x^*$ is a secretive envy-free solution. 

Say, under price vector $p^*$, the secretive agent selects room $k \in [n]$. Then, the property of the bijection $\pi_k: [n-1] \mapsto [n] \setminus \{k\}$ gives us $x^* \in \bigcap_{i \in [n-1]} C^i_{\pi_k(i)}$, i.e., for all agents $a \in \mathcal{A} \setminus \{ n \} = [n-1]$, the point $x^*$ is contained in the set $C^a_{\pi_k(a)}$. The definition of KKM covers (see Equation (\ref{definition:kkmcovers})) implies that the room $\pi_k(a)$ is a first choice of each agent $a \in \mathcal{A} \setminus \{n\}$.  

%\begin{align*}
%		v_a(\pi_k(a),h(x_{\pi_k(a)})) &\ge v_a(r,h(x_r)) \\ 
%		v_a(\pi_k(a),p_{\pi_k(a)}) &\ge v_a(r,p_r) \qquad \text{(By definition of function $h$)}
%	\end{align*}
	
Therefore, for any $k \in \mathcal{R} =[n]$, the allocation $(\pi_k, p^*)$ provides an envy-free solution. Overall, we get that $p^*$ is a secretive envy-free solution of the rent-division instance.
\end{proof}

\subsection{Algorithmic Result for Quasilinear Utilities}
Efficient algorithms for finding envy-free rent divisions (in the standard, non-secretive setting) have been developed in prior work under quasilinear utilities \cite{aragones1995derivation, Gal2017, klijn2000algorithm}. In this utility model each function $v_a(r, p_r)$ is of the form $B^a_r - p_r$, i.e., agent $a$'s utility for room $r$ when its rent is $p_r$
is equal to the base value, $B^a_r$, minus $p_r$. Note that Theorem $\ref{theorem:existence}$ applies to quasilinear utilities, since they are a subclass of bounded, continuous, and monotone decreasing utilities. 

%Note that $(\pi_k,p^k)$ is an envy free solution to the rent division instance $\langle \mathcal{A} \ \{n\} =[n-1], \mathcal{R} \setminus\{k\}, \{ v_a(r, \cdot) \}_{a\in \mathcal{A}\setminus \{n\}, r \in \mathcal{R} \setminus \{k\}} \  \rangle$ for every choice of room $k \in \mathcal{R}$ of the secretive agent. 

For a given rent-division instance $\mathcal{I} = \langle \mathcal{A}, \mathcal{R}, \{ v_a(r, \cdot) \}_{a \in \mathcal{A} \setminus \{n\} , r }  \rangle$ with a secretive agent and quasilinear utilities, throughout this subsection we will use $\mathcal{H} =  ( \left(\mathcal{A} \setminus \{n\} \right) \cup \mathcal{R}, \left( \mathcal{A}\setminus \{ n \} \right) \times \mathcal{R}) $ to denote the complete,  weighted bipartite graph (between agents and rooms) in which weight of each edge $(a, r)$ is equal to the base value $B^a_r$. Also, write $\mathcal{H}^k$ to denote the weighted, bipartite graph obtained by removing the $k$th vertex from the rooms' side in $\mathcal{H}$. 

Extending this notation, we will use $\overline{\mathcal{H}}$ to denote the analogous bipartite graph in the standard (non-secretive) rent-division context. Note that, in $\overline{\mathcal{H}}$, the two vertex sides are of size $n$, each, and the edges indecent on the $n$th vertex of the agents' side have weights $B^n_r$ for $r \in \mathcal{R}$. 
 
%Let $\mathcal{G}^k$ represent the graph $\mathcal{G}$ with vertex $k$ removed from the rooms' side. We will use this notation without preamble to denote the weighted bipartite graph between agents and rooms of a secretive rent division instance with quasilinear utilities across this subsection.\\

Next, using the above mentioned notation, we state the second welfare theorem. This result provides an important characterization of envy-free rent divisions in the standard (i.e., non-secretive) quasilinear setting. 
 
\begin{lemma}[Second Welfare Theorem; Mas-Colell et al. ({\cite[Chapter~16]{MWG+95microeconomic}})]%Don't know which paper to cite
	\label{lemma:walrasian}
Let $ \mathcal{I} = \langle \mathcal{A}, \mathcal{R}, \{ v_a(r, \cdot) \}_{a , r }  \rangle$ be a rent-division instance with quasilinear utilities (i.e., $ v_a(r,x) = B^a_r - x$ for all $a \in \mathcal{A}$ and $r \in \mathcal{R}$). Then, if $(\pi,p)$ is an envy free solution of $\mathcal{I}$ and $\sigma$ is a maximum-weight perfect matching in $\overline{\mathcal{H}}$, then $(\sigma, p)$ is also an envy-free solution of the instance. Furthermore, every agent receives the same utility under both the solutions.
\end{lemma}

The subsequent lemma provides an analogue of the second fundamental theorem for the secretive context. The lemma shows that if $\pi_k$ is a maximum weight matching in bipartite graph $\mathcal{H}^k$, for all $k \in [n]$, then for \emph{any} secretive envy-free solution $p^*$ and any (room) choice $k$ of the secretive agent, one can use $\pi_k$ to find an allocation that achieves envy freeness overall. In other words, maximum-weight matchings of bipartite graphs $\mathcal{H}^k$s suffice to certify that $p^*$ is indeed a secretive envy-free solution. 

% for \emph{any} secretive envy-free solution $p^*$ and any (room) choice $k$ of the secretive agent, one can always use a maximum weight matching of the bipartite graph $\mathcal{H}^k$ to find an allocation that achieves envy freeness. %In other words, given a price vector $p^*$ we can use maximum weight matchings of the bipartite graph $\mathcal{H}^k$ to certify that indeed $p^*$ is a secretive envy-free solution. 
We use this result to establish the correctness of Algorithm \ref{alg:secretiverent}. %, which is designed to compute secretive envy-free solutions under quasilinear utilities. 

%The following lemma provides a characterization of the associated bijections of a secretive envy free solution. This lemma may be thought of as either an analogue to the second welfare theorem while considering rent division instances with a secretive agent or as a generalization of the same. 
{
	\begin{algorithm}
		{
			%\RaggedRight
			%\noindent
			{\bf Input:} A rent division instance $\langle \mathcal{A}, \mathcal{R}, \{ v_a(r, \cdot) \}_{a\in \mathcal{A}\setminus \{n\}, r \in \mathcal{R}} \  \rangle$ with a secretive agent and quasilinear utilities $\{ v_a(r, \cdot) \}_{a\in \mathcal{A}\setminus \{n\}}$ (i.e., $ v_a(r,x) = B^a_r - x$ for all $a \in \mathcal{A} \setminus \{n\}$ and all $r$). \\
			%\RaggedRight{
			%\noindent
			{\bf Output:} A secretive envy-free solution $p^* \in \mathbb{R}^+_n$ and $n$ bijections $\{\pi_k\}_{k \in \mathcal{R}}$ such that for each  $k$, the bijection $\pi_k: \mathcal{A} \setminus \{n\} \mapsto \mathcal{R} \setminus \{k\}$ along with $p^*$ provides an envy-free solution. 			
			\caption{Secretive Envy-Free Rent Division Under Quasilinear Utilities}	
			\label{alg:secretiverent}
			\begin{algorithmic}[1]
				\STATE For all $k \in \mathcal{R}$, set $\pi_k$ to be a maximum-weight perfect matching in the bipartite graph $\mathcal{H}^k$.
				\STATE Set $p^* \in \mathbb{R}^n_+$ to be the solution of the following linear program \\
				\COMMENT{The goal here is to find a single price vector under which each $\pi_k$ is an envy-free allocation} 
				\begin{align*}
					\min_{x \in \mathbb{R}^n} & \ \  \sum_{r \in [n]} x_r \\
					\text{subject to } &  \ \ \ x_r \geq 0 \quad &\text{for all $r \in [n]$} \\ 
					&\ \ \  B^a_{\pi_k(a)} - x_{\pi_k(a)} \ge B^a_r - x_r \quad &\text{for all $a \in \mathcal{A} \setminus \{n\}$ and all $k, r \in \mathcal{R}$}
				\end{align*} 
				%\COMMENT{Since the utilities are quasilinear, $v_a(\pi_k(a),x_{\pi_k(a)}) \ge v_a(r,x_r)$ is a linear inequality in the decision variable $x$ as expanded above.}
				\RETURN price vector $p^*$ and the $n$ bijections $\{\pi_k\}_{k \in \mathcal{R}}$.
			\end{algorithmic}
		}
	\end{algorithm}
}

\begin{lemma}
	\label{lemma:maxwt}
Let $\mathcal{I} = \langle \mathcal{A}, \mathcal{R}, \{ v_a(r, \cdot) \}_{a \in \mathcal{A} \setminus \{n\} , r }  \rangle$ be a rent-division instance with a secretive agent and quasilinear utilities (i.e., $ v_a(r,x) = B^a_r - x$ for all $a \in \mathcal{A} \setminus \{n\}$ and $r \in \mathcal{R}$). In addition, for all $k \in [n]$, let $\pi_k$ be a maximum-weight perfect matching in the bipartite graph $\mathcal{H}^k$ and $p^* \in \mathbb{R}^n_+$ be any secretive envy-free solution (price vector) of $\mathcal{I}$. Then, $(\pi_k, p^*)$---for all $k \in [n]$---is an envy-free solution of $\mathcal{I}$, i.e., $v_a(\pi_k(a), p_{\pi_k(a)}) \geq v_a(r,p_r)$ for all $k \in [n]$, $a\in \mathcal{A} \setminus \{n\}$ and $r \in \mathcal{R}$.
%Then, under any secretive envy-free solution (price vector) $p^* \in \mathbb{R}^n_+$, every bijection $\pi_k$ provides an envy-free allocation, i.e., for all $k \in [n]$, $(\pi_k, p^*)$ is an envy-free solution for agents in $ \mathcal{A} \setminus \{n\}$.%the $n$ bijections $\{\sigma_k\}_{k \in \mathcal{R}}$ can be used to obtain the required envy free allocations. 
% there exists a price vector $p^*$ such that the $n$ bijections $\{\sigma_k\}_{k \in \mathcal{R}}$ along with $p^*$ constitute a secretive envy-free solution of $\mathcal{I}$.  i.e., for all agents $a$ in $\mathcal{A} \setminus \{n\}$, all rooms $r$ in $\mathcal{R}$ and each choice $k \in \mathcal{R}$ of the secretive agent: 
\end{lemma}

\begin{proof}
Theorem~\ref{theorem:existence} implies that instance the $\mathcal{I}$ (with quasilinear utilities) admits a secretive envy-free solution $p^*$. Therefore, by definition of a secretive solution, there exist $n$ bijections, $\phi_1, \phi_2, \ldots, \phi_n$, that provide that an envy-free allocation for every possible choice (room selection) of the secretive agent. Specifically, for every choice $k \in \mathcal{R}$, all agents $a \in \mathcal{A} \setminus \{n\}$, and all rooms $r \in \mathcal{R}$ we have  $v_a(\phi_k(a),p_{\phi_k(a)}) \geq v_a(r, p_r) $.

For all $k \in \mathcal{R}$, write $p^k \in \mathbb{R}^{n-1}$ to denote the restriction of the price vector $p$ to the rooms $\mathcal{R} \setminus \{k\}$. Note that $(\phi_k, p^k)$ is an envy-free solution of the (non-secretive) rent-division instance $\mathcal{I}^k = \langle \mathcal{A} \setminus \{n\} =[n-1], \mathcal{R} \setminus\{k\}, \{ v_a(r, \cdot) \}_{a\in \mathcal{A}\setminus \{n\}, r \in \mathcal{R} \setminus \{k\}} \rangle$. 

Given that all the utilities in $\mathcal{I}^k$ are quasilinear, we can apply the Second Welfare Theorem (Lemma \ref{lemma:walrasian}) to obtain that $(\pi_k, p^k)$ is also an envy-free solution of this instance; recall that $\pi_k$ is a maximum-weight matching in $\mathcal{H}^k$. In addition, Lemma \ref{lemma:walrasian} guarantees that agent $a$ obtains the same utility under both the solutions $(\phi_k, p^k)$ and $(\pi_k, p^k)$, i.e., $v_a(\phi_k(a), p_{\phi_k(a)}) = v_a(\pi_k(a), p_{\pi_k(a)})$. Therefore, $v_a(\pi_k(a), p_{\pi_k(a)}) \geq v_a(r, p_r)$ for all rooms $r \in \mathcal{R}$. 

In other words, the bijections $\{\pi_k\}_{k \in \mathcal{R}}$ also provide an envy-free allocation for every possible choice of the secretive agent and satisfy the required inequality $v_a(\pi_k(a), p_{\pi_k(a)}) \geq v_a(r,p_r)$ for all $k \in [n]$, $a\in \mathcal{A} \setminus \{n\}$ and $r \in \mathcal{R}$.
\end{proof}

Theorem~\ref{theorem:secretive-ef} states the correctness and time complexity of Algorithm~\ref{alg:secretiverent}.

%The above result, in conjunction with \cref{lemma:despised}, is critical to our algorithmic results regarding secretive $\EFone$ allocations of indivisible goods.

\TheoremRent*

\begin{proof}
	%%%First we will prove that Algorithm~\ref{alg:secretiverent} indeed computes a secretive envy-free solution of the given instance $\mathcal{I}$.
	Theorem $\ref{theorem:existence}$ ensures that the given instance $\mathcal{I}$ admits a secretive envy-free solution $p \in \mathbb{R}^n$. We can assume that $p$ is componentwise nonnegative--uniform, additive shifts of a price vector maintain envy freeness under quasilinear utilities. In addition, using Lemma~\ref{lemma:maxwt}, for each $\pi_k$ (a maximum-weight matching in $\mathcal{H}^k$) we have $v_a(\pi_k(a),p_{\pi_k(a)}) \ge v_a(r,p_r) $, i.e., $\pi_k$s provide envy-free allocations for every choice $k$ of the secretive agent. 
	
	Therefore, $p$ certifies the feasibility of the  linear program (LP) formulated in Algorithm~\ref{alg:secretiverent}. Since this LP is also bounded, the algorithm will necessarily find a price vector $p^*$ which (paired with the bijections $\{\pi_k \}_{k \in \mathcal{R}}$) explicitly satisfies the definition of a secretive envy-free solution. This proves the correctness of Algorithm~\ref{alg:secretiverent}. 
	Finally, note that Algorithm~\ref{alg:secretiverent} entails finding $n$ maximum-weight matchings and solving a linear program with a polynomial number of variables and constraints. Hence, the algorithm runs in polynomial time and the stated claim follows. \\ 
\end{proof}

%\cref{alg:secretiverent} uses the characterization of the bijections associated with a secretive envy free solution to find a valid collection of bijections in polynomial time. Then, the linear program associated with envy free solutions of a quasilinear rent division instance is solved to generate the secretive envy free solution. Although we do not know much about the structure of price vectors satisfying these conditions, the proof of existence given by \cref{theorem:existence} is sufficient to show that the linear program is feasible. An interesting open problem is that of finding a combinatorial construction of this price vector, along the lines of Aragones ~\cite{} for the classical envy free rent division problem with quasilinear utlilities.

%In course of proving the correctness of \cref{alg:secretiverent}, we proved the following result characterizing the associated set of bijections of a secretive envy free solution to a rent division instance with quasilinear utilities.

We conclude this section by stating and proving a combinatorial lemma, which will be essential in establishing the $\EFone$ results. The lemma asserts that for with respect to maximum weight matchings $\pi_k$s (in graphs $\mathcal{H}^k$s) there will always be a room that is ``universally despised'' by all the agents and across all the $n$ bijections $\pi_k$s.  
 
%Following up on this existence result, we state and prove an interesting lemma that is critical for our algorithmic result regarding secretive $\EFone$ allocations.  
\begin{lemma}
\label{lemma:despised}
Let $\mathcal{I} = \langle \mathcal{A}, \mathcal{R}, \{ v_a(r, \cdot) \}_{a \in \mathcal{A} \setminus \{n\} , r }  \rangle$ be a rent-division instance with a secretive agent and quasilinear utilities (i.e., $ v_a(r,x) = B^a_r - x$ for all $a \in \mathcal{A} \setminus \{n\}$ and $r \in \mathcal{R}$). In addition, for all $k \in [n]$, let $\pi_k$ be a maximum weight perfect matching in the bipartite graph $\mathcal{H}^k$. Then, there always exists a room $\rho \in [n]$ such that $B^a_{\pi_k(a)} \geq B^a_\rho$, for all agents $a \in \mathcal{A} \setminus \{n\}$ and all bijections $\pi_k$, with $k \in [n]$. % \in \{\pi_k\}_{k \in [n]}$. 
	
%Given any secretive envy free solution $p \in \mathbb{R}^n_+$ along with an associated of bijections $\{\pi_k\}_{k \in [n]}$ of a rent division instance with a secretive agent, represented by the tuple $\langle \mathcal{A} =[n], \mathcal{R}=[n], \{ v_a(r, \cdot) \}_{a\in \mathcal{A}\setminus \{n\}, r \in \mathcal{R}} \  \rangle$, where the utilities functions $v_a(r,z)$s are quasilinear (of the form $B^a_r - z$), there always exists a room $\rho \in [n]$ with the following property: for all agents $a \in \mathcal{A} \setminus \{n\}$ and all bijections $\pi \in \{\pi_k\}_{k \in [n]}$: $B^a_{\pi(a)} \ge B^a_\rho$ i.e. the room $\rho$ is "universally despised" across the $n$ bijections.
\end{lemma}
\begin{proof}
Write $p$ to denote a secretive envy-free solution of the quasilinear instance $\mathcal{I}$; we know that such a price vector exists via Theorem~\ref{theorem:existence}. Consider a room $\rho \in \mathcal{R}$ with the smallest rent under $p$: $\rho \in \argmin_{r \in \mathcal{R}} p_r$. Since $p$ is a secretive envy-free solution, for any choice $k \in \mathcal{R}$ of the secretive agent, $\pi_k$ provides an envy-free allocation (Lemma \ref{lemma:maxwt}). In particular, at the imposed prices, no agent $a$ strongly prefers room $\rho$ to the room allocated to her, i.e., to $\pi_k(a)$. Therefore, 
\begin{align*}
v_a(\pi_k(a), p_{\pi_k(a)}) & \geq v_a(\rho, p_\rho)   & \text{(($\pi_k,p$) is an envy-free solution)}  \\
B^a_{\pi_k(a)} - p_{\pi_k(a)} & \geq B^a_\rho - p_\rho & \text{(utilities are quasilinear)} \\
B^a_{\pi_k(a)} & \geq B^a_\rho  & \text{(since $p_\rho \leq p_{\pi_k(a)}$)} %\in \argmin_{r \in \mathcal{R}} p_r$)} 		 
\end{align*}
\end{proof}

\section{Envy-Free Division with a Secretive Agent}
\label{section:envy}
In Section~\ref{section:EF1}, we address indivisible goods and show that a secretive $\EFone$ allocation always exists and can be computed efficiently, as long as the valuations of the non-secretive agents are monotone increasing. 

Section~\ref{section:epsEF} provides results for divisible goods (i.e., cake cutting). In particular, we reduce the problem of finding a secretive $\epsEF$ solution (in the cake-cutting setup) to that of computing a secretive $\EFone$ allocation (over indivisible goods). Therefore, using the results developed in Section~\ref{section:EF1}, we obtain existential and algorithmic results for secretive $\epsEF$ division.

\subsection{Secretive EF1 Allocations of Indivisible Goods}
\label{section:EF1}
We begin by stating notation which will be required for detailing the algorithm that efficiently finds a secretive $\EFone$ allocation (Algorithm~\ref{alg:ef1}). Consider a fair division instance $\mathcal{I} = \langle \mathcal{A} =[n], \mathcal{G}=[m], \{ v_a \}_{a\in \mathcal{A}\setminus \{n\}} \  \rangle$ with a secretive agent and indivisible goods $\mathcal{G}$. Let $ \mathcal{P} = (P_1,P_2,...P_n)$ be an $n$-partition of a subset of goods $ \mathcal{G}' \subseteq \mathcal{G}$; we will refer to such an $n$-partition $\mathcal{P}$ as a partial allocation. Note that $\mathcal{G}'= \cup_i P_i$.

Given any partial allocation $\mathcal{P}$, write $\mathcal{H}_{\mathcal{P}}:= \left( (\mathcal{A} \setminus \{ n \}) \cup \{P_i\}_i,  (\mathcal{A} \setminus \{ n \}) \times \{P_i \}_i \right) $ to denote the weighted, bipartite graph whose bipartition is composed of the set of agents $\mathcal{A} \setminus \{ n \}$ and the bundles $\{P_1, \ldots P_n \}$. Here, an edge $(a,P_i) \in  (\mathcal{A} \setminus \{ n \}) \times \{P_1, \ldots, P_n \}$ is said to be satisfy the $\EFone$ property iff for all bundles $P_j$ in $\mathcal{P}$ there exists a good $g \in P_j$ such that $v_a(P_i) \geq v_a(P_j \setminus \{g\})$.

Write $\mathcal{E}_\mathcal{P}$ to denote the set of edges that satisfy the $\EFone$ property and $\mathcal{E}^c_\mathcal{P}$ to denote the set of edges that do not, $\mathcal{E}^c_\mathcal{P} = \left((\mathcal{A} \setminus \{ n \}) \times \{P_i \}_i \right) \setminus \mathcal{E}_\mathcal{P}$. In $\mathcal{H}_\mathcal{P}$, the weight of each edge $(a,P_i)$ that satisfies the $\EFone$ property is set to be $v_a(P_i)$ and all edges $(b, P_j) \in \mathcal{E}^c_\mathcal{P}$ are assigned a weight equal to $-m \left( \max_{a \in \mathcal{A} \setminus \{n\}} v_a(\mathcal{G})\right)$, where $m$ is the total number of goods. 

Finally, we will use ${\mathcal{H}}^k_\mathcal{P}$ to denote the bipartite graph obtained by removing vertex $P_k$ from the graph ${\mathcal{H}}_\mathcal{P}$. Note that the partition $\mathcal{P}$ is a secretive $\EFone$ allocation if and only if, for every possible (choice of the secretive agent) $k \in [n]$, there exists a perfect matching $\pi_k$ in ${\mathcal{H}}^k_\mathcal{P}$ that contains only $\EFone$ edges, i.e., satisfies $\pi_k \subseteq \mathcal{E}_\mathcal{P}$.\footnote{Here, for ease of presentation, we overload notation and use $\pi_k$ to represent both a bijection and a matching (subset of edges).} Here, the choice of edge weights ensures that if there is a perfect matching in $\mathcal{H}^k_{\mathcal{P}}$ consisting solely of $\EFone$ edges, then any maximum weight matching will also be composed entirely of $\EFone$ edges. 

Next, we will use these constructs to describe Algorithm \ref{alg:ef1} and prove its correctness in Theorem \ref{theorem:ef1secretive}. A key argument in the proof is an application of Lemma \ref{lemma:univ-disp} to show that  Step \ref{step:sink} of Algorithm \ref{alg:ef1} always succeeds. In particular, Lemma \ref{lemma:univ-disp}  guarantees the existence of a ``universally-despised'' bundle.

{
	\begin{algorithm}
		{
			%\RaggedRight
			%\noindent
				{\bf Input:} A fair division instance $\mathcal{I} = \langle \mathcal{A} , \mathcal{G}, \{ v_a \}_{a\in \mathcal{A}\setminus \{n\}} \  \rangle$ with a secretive agent, $m$ indivisible goods along with nonnegative and monotone valuations (set functions) $v_a: 2^{[m]}\rightarrow \mathbb{R}_+$ for agents $a \in \mathcal{A} \setminus \{n \}$ \\ 
				%\RaggedRight{
				%\noindent
				{\bf Output:} An allocation $(P_1, \ldots, P_n)$ and $n$ bijections $\{\pi_k\}_{k \in [n]}$ such that, for each choice $k$ of the secretive agent, $\pi_k: \mathcal{A} \setminus \{n\} \mapsto \mathcal[n] \setminus \{k\}$ induces an $\EFone$ allocation.
		 \caption{Computation of Secretive $\EFone$ Allocations}	
			\label{alg:ef1}
			\begin{algorithmic}[1]
				\STATE Initialize $t \leftarrow 0$ and set $P^t_a = \emptyset$ for $1 \leq a \leq n$, i.e., partial allocation $\mathcal{P}^0 = (\emptyset, \ldots, \emptyset)$.		
%				\STATE For each $k \in [n]$, initialize $\pi^t_k: \mathcal{A} \setminus \{n\} \mapsto [n] \setminus \{k\}$ to be an arbitrary bijection.
				\WHILE{$\mathcal{G} \neq \emptyset$}
				\STATE For partial allocation $\mathcal{P}^t=(P^t_1, \ldots, P^t_n)$, construct the bipartite graph ${\mathcal{H}}_{\mathcal{P}^t}$
				\STATE For each $k \in [n]$, set $\pi_k$ to be a maximum-weight perfect matching in the graph ${\mathcal{H}}^k_{\mathcal{P}^t}$ (i.e., in the graph ${\mathcal{H}}_{\mathcal{P}^t}$ with vertex $P^t_k$ removed).
				\STATE \label{step:sink} Find a bundle $P^t_\rho$ such that for all $k \in [n]$ and all $a \in \mathcal{A} \setminus \{n\}$ we have  $v_a(P^t_{\pi_k(a)}) \geq v_a(P^t_{\rho})$ 
				\COMMENT{We will prove that such a ``universally-despised'' bundle always exists and can be found efficiently}
				\STATE Select an arbitrary good $g \in \mathcal{G}$ and set $P^{t+1}_\rho = P^t_\rho \cup \{g\} $
				\STATE  For all $i \in [n] \setminus \{ \rho \}$ set $P^{t+1}_i = P^t_i$. Update $t \leftarrow t +1$ and $\mathcal{G} \leftarrow \mathcal{G} \setminus \{g\}$
				\ENDWHILE
				\RETURN Partition $P^t = (P^t_1,P^t_2,...,P^t_n)$ and the $n$ bijections $\{\pi_k\}_{k \in [n]}$
			\end{algorithmic}
		}
	\end{algorithm}
}

\begin{restatable}{lemma}{LemmaUnivdesp}
	\label{lemma:univ-disp}
	Let $\mathcal{P}=(P_1, \ldots, P_n)$ be a partial allocation of the indivisible goods in an instance ${\mathcal{I}} = \langle \mathcal{A}, \mathcal{G}, \{ v_a \}_{a\in \mathcal{A}\setminus \{n\}} \  \rangle$ and let $\pi_k$ be a maximin weight perfect matching in the bipartite graph ${\mathcal{H}}^k_\mathcal{P}$, for each $k \in [n]$. If the matching $\pi_k$s are composed entirely of $\EFone$ edges ($\pi_k \subseteq \mathcal{E}_\mathcal{P}$), then there exists an index (bundle) $\rho \in [n]$ such that $v_a(P_{\pi_k(a)}) \geq v_a(P_\rho)$ for all $a \in \mathcal{A} \setminus\{n\}$ and all $k \in [n]$. 
\end{restatable}

\begin{proof}
	From the given instance $\mathcal{I}$ and partial allocation $\mathcal{P}=(P_1, \ldots, P_n)$, we construct a rent-division instance $\widehat{\mathcal{I}} = \langle \mathcal{A}, \mathcal{R}=[n], \{ v_a(r, \cdot) \}_{a\in \mathcal{A}\setminus \{n\}, r \in \mathcal{R}} \rangle$ in which the $i$th room corresponds to the $i$th bundle $P_i$. The utilities in $\widehat{\mathcal{I}}$ are set to be quasilinear, $v_a(i, z) :=B^a_i - z$. Here, $B^a_i$ (the base value of room $i$ for agent $a$) is assigned to the weight of the edge $(a, P_i)$ in the bipartite graph $\mathcal{H}_{\mathcal{P}}$. That is, if edge $(a, P_i)$ satisfies the $\EFone$ property then $B^a_i = v_a(P_i)$. Otherwise, if $(b, P_j)$ is not an $\EFone$ edge, then $B^b_j = -m  \left( \max_{a \in \mathcal{A} \setminus \{n\}} v_a(\mathcal{G})\right)$. 
	
	An application of Lemma~\ref{lemma:despised} over rent-division instance $\widehat{\mathcal{I}}$ shows that there exists a room $\rho \in [n]$ such that---for all agents $a \in \mathcal{A} \setminus \{n\}$ and maximum weight matchings $\pi_k$---we have  $B^a_{\pi_k(a)} \geq B^a_\rho$.\footnote{By construction of $\widehat{\mathcal{I}}$, the bipartite graph $\mathcal{H}^k$ considered in Lemma~\ref{lemma:despised} is identical to $\mathcal{H}^k_{\mathcal{P}}$. Hence, we can apply the lemma to $\pi_k$s, the maximum wight perfect matchings of $\mathcal{H}^k_{\mathcal{P}}$s.}
	
	%Note that in the bipartite graph $\mathcal{H}^k_{\mathcal{P}}$ the edges in that do not satisfy that $\EFone$ property (i.e., edges  in the set $\mathcal{E}^c_{\mathcal{P}}$) are assigned a negative weight. The weight is low enough to ensure that if $\mathcal{H}^k_{\mathcal{P}}$ admits a perfect matching composed entirely of $\EFone$ edges, then a maximum weight matching, $\pi_k$, in $\mathcal{H}^k_{\mathcal{P}}$ will satisfy $\pi_k \subseteq  \mathcal{E}_{\mathcal{P}}$. 
	
	Using the fact that every edge in the matching $\pi_k$ is an $\EFone$ edge, we get $B^a_{\pi_k(a)} = v_a(P_{\pi_k(a)}) $ for all agents $a \in \mathcal{A} \setminus \{ n \}$. 
	
	Now, if for an agent $a$ edge $(a, P_\rho)$ satisfies the $\EFone$ property then $B^a_\rho = v_a(P_\rho)$ and we get the desired inequality $v_a(P_{\pi_k(a)}) \geq v_a(P_\rho)$. Otherwise, if edge $(a, P_\rho)$ is not an $\EFone$ edge, then there must exist a bundle $P_j$ such that for all $g' \in P_j$ the inequality $v_a(P_\rho) < v_a(P_j \setminus \{ g'\})$ holds. Again, using the fact that $(a, P_{\pi_k(a)})$ is an $\EFone$ edge, we get that there exists $\tilde{g} \in P_j$ for which $v_a(P_{\pi_k(a)}) \geq v_a(P_j \setminus \{ \tilde{g}\})$. Hence, the desired inequality $v_a(P_{\pi_k(a)}) \geq v_a(P_\rho) $ holds in all cases.
\end{proof}

\TheoremEFone*
%Proof Structure - By Induction+Initial allocation
%In each round, existence of despised bundle - by invoking the lemma
% Matchings composed only of EF1 edges- existence and rewighting
% Argue existence of new set of ef1 bijections
\begin{proof}
We will use an inductive argument to establish the correctness of Algorithm~\ref{alg:ef1}. Write $\mathcal{P}^t$ to denote the partial allocation considered by Algorithm~\ref{alg:ef1} in the $t$th iteration of the while-loop. We will show that the maximum-weight matching matchings computed by algorithm during any iteration $t$, say $\pi_k$s, satisfy $\pi_k \subseteq \mathcal{E}_{\mathcal{P}^t}$, i.e., in all iterations, the computed matchings are composed entirely of $\EFone$ edges. In conjunction, we will prove that there exits a bundle $P^t_\rho$ such that the inequality $v_a(P^t_{\pi_k(a)}) \ge v_a(P^t_\rho)$ holds for all agents $a \in \mathcal{A} \setminus \{n\}$ and all matchings $\pi_k$.  

Both of these conditions hold when $t=0$, since $P^0_i = \emptyset$ for all $i \in [n]$. This gives us the base case for induction. Now, say---via the induction hypothesis---that the conditions hold for the $(t-1)$th iteration. In particular, if $\widehat{\pi}_k$s are the maximum-weight matchings considered in the $(t-1)$th iteration, then we have $\widehat{\pi}_k \subseteq \mathcal{E}_{\mathcal{P}^{t-1}} $ and $v_a(P^{t-1}_{\widehat{\pi}_k(a)}) \geq v_a(P^{t-1}_{\widehat{\rho}})$ for some room $\widehat{\rho}$. 

By construction,  $P^t_{\widehat{\rho}} = P^{t-1}_{\widehat{\rho}} \cup \{ g \}$ and $P^t_i = P^{t-1}_i$ for all $i \neq \widehat{\rho}$. Note that, even after this update, the edges in matchings $\widehat{\pi}_k$s continue to satisfy the $\EFone$ property: if $(a, P^{t-1}_i)$ is an edge in $\widehat{\pi}_k$ (i.e., $i = \widehat{\pi}_k(a)$), then $v_a(P^{t-1}_i) \geq v_a(P^{t-1}_{\widehat{\rho}})$. Therefore, $v_a(P^{t}_i) \geq v_a(P^{t}_{\widehat{\rho}} \setminus \{ g \})$. Since no other bundle receives a good, the $i$th bundle satisfies the $\EFone$ property for $a$ in the new partial allocation $\mathcal{P}^t= (P^t_1, \ldots, P^t_n)$ as well. 

This implies that there exists a perfect matching in the bipartite graph $\mathcal{H}^k_{\mathcal{P}^t}$ (specifically, $\widehat{\pi}_k$) which is entirely composed on $\EFone$ edges. The weight of all the edges in $\mathcal{E}^c_{\mathcal{P}^t}$ is low (negative) enough to ensure that if $\mathcal{H}^k_{\mathcal{P}^t}$ admits a perfect matching composed entirely of $\EFone$ edges, then a maximum-weight matching, $\pi_k$, in $\mathcal{H}^k_{\mathcal{P}^t}$ will satisfy $\pi_k \subseteq  \mathcal{E}_{\mathcal{P}^t}$. 

Furthermore, Lemma~\ref{lemma:univ-disp} (applied to $\mathcal{P}^t$) implies that there exists a bundle $P^t_\rho$ which satisfies the desired inequalities $v_a(P^t_{\pi_k(a)}) \ge v_a(P^t_\rho)$. This concludes the inductive argument and shows that Step~\ref{step:sink} will necessarily succeed in finding bundle $P^t_\rho$.  Also, note that with matchings $\pi_k$s in hand, an exhaustive search can be performed to efficiently find $P^t_\rho$. Hence, the algorithm runs in polynomial time. 
%The algorithm terminates in polynomial time: it iterates $m$ times (assigning one indivisible good in each run of the while-loop) and each iteration entails finding $n$ maximum weight perfect matchings along with searching for a bundle $P_\rho$. 

Note that if $\mathcal{P}$ is the final partition and $\pi_k$s are the corresponding (returned) matchings, then we know that $\pi_k$s are composed entirely of $\EFone$ (with respect to bundles in $\mathcal{P}$) edges. Therefore, $\mathcal{P}$ is a secretive $\EFone$ allocation and the claim follows.  
\end{proof}

\subsection{Secretive $\epsEF$ Cake Cutting}
\label{section:epsEF}

In this subsection, we present an algorithm to find a secretive $\epsEF$ solutions of cake-cutting instances. This algorithm is based on a reduction of Lipton et. al. \cite{lipton2004approximately}, which transforms the problem of (approximately) envy free cake-cutting to that of finding a secretive $\EFone$ allocation of indivisible goods. 

\TheoremEFCake*

\begin{proof}
We query the agents' valuations to divide the cake into continuous pieces such that each piece is of value at most $\varepsilon$ for any agent in $\mathcal{A} \setminus \{n\}$. As the valuation functions are additive, there are at most $n/\varepsilon$ such pieces. Considering these pieces as indivisible goods, we find a secretive $\EFone$ allocation $\mathcal{P}$ using \cref{theorem:ef1secretive}. Note that the envy in such an allocation is upper bounded by the largest value for a single piece, which is at most $\varepsilon$. Therefore, $\mathcal{P}$ is a secretive $\epsEF$ solution  of instance given cake-cutting instance $\mathcal{I}$.
\end{proof}

\section{Proportional and Maximin Fair Division with a Secretive Agent}
\label{section:propmms}
This section considers two settings: proportional division of a cake (a divisible good) and maximin fair division of indivisible goods. In both of these cases we show that, even in presence of a secretive agent, we can establish existential and algorithmic results that are essentially similar to ones obtained in the standard (non-secretive) settings.  

% cake cutting for the setting of divisible goods and maximin allocations for the setting of indivisible goods. In both these settings, we generalize the idea of cut-and-choose protocol by extending it to $n$ agents.

\subsection{Proportional Cake Cutting with a Secretive Agent}
\label{subsection:propcake}

This subsection presents a polynomial-time algorithm for computing a secretive proportional solution of a given cake; the definition of this solution concept appears in Section \ref{section:problems}. In particular, we show that the Dubins-Spanier moving knife procedure \cite{dubins1961cut} when executed among the first $n-1$ agents, with threshold set to $1/n$, finds such a solution. 
%

%This procedure can be interpreted as a generalization of cut-and-choose protocol, where $(n-1)$ agents together make cuts on the cake and let the secretive agent choose first. 

\TheoremPropcake*

\begin{proof}
The Dubins-Spanier moving knife procedure can be executed in polynomial time under the Robertson-Webb query model. Specifically, for each agent $a \in \mathcal{A}\setminus\{n \}$, we query Cut($a, [0,1],1/n$) and select an agent $a_1$ with the (inclusion wise) smallest cut, i.e., agent $a_1 = \argmin_{a \in \mathcal{A} \setminus\{ n\}} \text{Cut}(a, [0,1],1/n)$ makes the first cut on the cake. We continue the same procedure over the remaining cake---with agent $a_1$ removed consideration---until each of the $n-1$ agents makes a cut.  The fact that the agents' valuations are additive implies that every agent will end up making a cut. Also, note that at the end of the procedure the cake gets partitioned into exactly $n$ pieces. 

%That is, we can efficiently simulate the following procedure: we move a knife is moved from one end of the cake to the other and the cake is cut as soon as, for any agent $a \in \mathcal{A} \setminus \{ n \}$ (who does not have a piece yet), the value of the piece to the left of the knife is $1/n$ times the total value of the cake. This piece is allocated to agent $a$ and she is removed from consideration. The last agent gets the remainder of the cake. 

We reindex the agents based on the order in which they made a cut. That is, agent $ i \in [n-1]$ makes a cut in the $i$th round and reserves, say, the piece $P_i$. The procedure ensures that $v_i(P_i) = 1/n$ for each $i$.\footnote{Recall that the valuations are normalized to satisfy $v_a([0,1]) = 1$ for all $a\in \mathcal{A} \setminus \{ n \} $.} Write $P_n$ to denote the last (leftover) piece. In addition, note that the procedure ensures that for any agent $i \in [n-1]$ the proportional value is not exceeded by any piece with a lower index, $v_i(P_k) \leq 1/n$ for all $k < i$. 

Hence, for each $i$ we have $\sum_{k: k>i} v_i(P_k) \geq 1 - 1/n - (i-1)/n = (n-i)/n$. An averaging argument guarantees the existence of a piece $P_j$ such that $j>i$ and $v_i(P_j) \geq 1/n$. We select an index $j$ that satisfies this inequality and set $\sigma(i) =j$. The mapping $\sigma$ can be thought of as a backup-allocation scheme: for every agent $i$, both the pieces $P_i$ and $P_{\sigma(i)}$ guarantee a proportional share. Since $\sigma(i)>i$ for each agent $i \in [n-1]$, using Lemma~\ref{lemma:backup}, we can construct $n$ bijections $\{\pi_k\}_{k \in [n]}$ that guarantee a proportional share for every possible choice $P_k$ of the secretive agent. In other words, the computed partition $(P_1, \ldots, P_n)$ is a secretive proportional solution. This completes the proof of the theorem. 
\end{proof}	

\subsection{$\MMS$ Allocation of Indivisible Goods with a Secretive Agent}
\label{section:mms}
This subsection presents a polynomial-time algorithm (Algorithm \ref{alg:mms}) for computing secretive (approximate) $\MMS$ allocations under monotone, submodular valuations. 

For a monotone increasing, submodular valuation $v: 2^G \mapsto \mathbb{R}_+$, over the set of indivisible goods $G$, we  define a thresholding procedure to obtain a surrogate valuation (set function) $\widehat{v}: 2^G \mapsto \mathbb{R}_+$ whose marginals are upper bounded. The function $\widehat{v}$ will continue to be monotone and submodular, and this construct will be used in Algorithm~\ref{alg:mms}. Formally, given threshold $\kappa >0$, a specific good $\widehat{g} \in \mathcal{G}$, and a monotone, submodular valuation $v: 2^G \mapsto \mathbb{R}_+$ we define a set function $\widehat{v}: 2^G \mapsto \mathbb{R}_+$ as follows   

%Given an instance $\mathcal{I} = \langle \mathcal{A}, \mathcal{G}, \{ v_a \}_{a\in \mathcal{A}\setminus \{n\}} \  \rangle$ with a secretive agent, indivisible goods $\mathcal{G}=[m]$, and monotone, submodular valuations $v_a: 2^\mathcal{G} \mapsto \mathbb{R}_+$, we define a thresholding procedure that (under appropriate conditions) provides another set of valuations $\widehat{v}_a: 2^\mathcal{G} \mapsto \mathbb{R}_+$ with bounded marginals. This procedure maintains monotonicity and submodularity as well as approximately preserves the maximin share values of the agents. This construct will be useful in Algorithm~\ref{alg:mms}. Formally, given threshold $\kappa >0$, a specific good $g \in \mathcal{G}$, and a monotone, submodular valuation $v: 2^\mathcal{G} \mapsto \mathbb{R}_+$ we define the set function $\widehat{v}: 2^\mathcal{G} \mapsto \mathbb{R}_+$ as follows

\begin{align}
\label{eqn:normalize}
\widehat{v}(S) := 
\begin{cases}
v(S) & \text{ if } \widehat{g} \notin S \\
v(S\setminus\{ \widehat{g} \}) + \min \left\{ \kappa,   v_{S\setminus \{\widehat{g} \}} (\widehat{g})  \right\} & \text{ else, if } \widehat{g} \in S
\end{cases}
\end{align}

Here $v_{S\setminus \{\widehat{g} \}} (\widehat{g})$ denotes the marginal value of good $\widehat{g}$ with respect to the set $S\setminus \{\widehat{g} \}$, i.e., $v_{S\setminus \{\widehat{g} \}} (\widehat{g}) = v (S \cup \{\widehat{g} \} ) - v (S\setminus \{\widehat{g} \})$.
The next lemma provides relevant properties of $\widehat{v}$ and its proof appears in Appendix~\ref{section:submodularmono}.

{
	\begin{algorithm}[ht!]
		{
			%\RaggedRight
			%\noindent
			{\bf Input:} An instance $\mathcal{I} = \langle \mathcal{A} =[n], \mathcal{G}=[m], \{ v_a \}_{a\in \mathcal{A}\setminus \{n\}} \  \rangle$ and thresholds $\tau_a$ for each $a \in \mathcal{A} \setminus \{n\}$. The nonnegative, monotone, submodular valuations, $v_a$s, are specified via an oracle. 	\\
			%\RaggedRight{
			%\noindent
			{\bf Output:} An allocation $(P_1, \ldots, P_n)$ and a mapping $\sigma:[n-1] \mapsto [n]$ such that, for all $a \in \mathcal{A} \setminus \{n\}$, we have $v_a(P_a) \geq \tau_a/19$, $v_a(P_{\sigma(a)}) \geq \tau_a/19$, and $\sigma(a) > a$.
			
			\caption{Computation of Approximate Secretive $\mms$ Allocations}	
			\label{alg:mms}
			\begin{algorithmic}[1]
				%\STATE Using Theorem \ref{thm:Apx-mms-value}, compute approximate maximin share $\mu'_i \in [\frac{1}{9}\mu_i,\mu_i]$ for each agent $i$.
				\STATE Initialize set of agents $A=[n-1]$ and set of goods $G=[m]$.
				\WHILE{there exist agent $a \in A$ and goods $g, g' \in G$ such that $v_a(g) \geq  \frac{\tau_a}{19}$ and $v_a(g') \geq \frac{\tau_a}{19}$}
				\STATE Allocate $P_a \leftarrow\{g\}$, and update $A \leftarrow A \setminus\{a\}$ and $ G \leftarrow G\setminus\{g\}$.
				\ENDWHILE \\
				\COMMENT{At this point, for each agent $a \in A $ there is at most one good $g \in G$ which satisfies $v_a(g) \geq \frac{\tau_a}{19}$}
				\STATE \label{step:aprime} Set $A' = \{ a \in A \mid \text{ there exists } g_a \in G  \text{ with } v_a(g_a) \geq \frac{\tau_a}{19} \}$.
				\STATE For set of goods $G$ and each $a \in A'$, obtain $\widehat{v}_a: 2^G \mapsto \mathbb{R}_+$ by applying (\ref{eqn:normalize}) with $\widehat{g} = g_a$ and $\kappa = \frac{\tau_a}{19}$ 
				\STATE For each $a \in A \setminus A'$, set $\widehat{v}_a = v_a$ \\ \COMMENT{In the subsequent steps we only require oracle access to $\widehat{v}_a$ restricted to $G$}
				\STATE \label{step:subinstance} Find a $1/3$-$\MMS$ allocation $(Q_1, Q_2, \ldots, Q_{|A|})$ for the the instance $\langle A, G, \{\widehat{v}_a\}_{a \in A} \rangle$ . \COMMENT{This can be accomplished using the polynomial-time algorithm from \cite{ghodsi2018fair}}
				\IF{for an agent $a \in A$ we have $v_a(Q_a) < \frac{3}{19} \tau_a$ }
				\STATE \label{step:hightau} Flag agent $a$ and exit. \COMMENT{We will show that this condition is executed only if $\tau_a > \mu_a$.} 
				\ENDIF
				\STATE For each $a \in A$, partition each $Q_a$ into two subsets $Q'_a$ and $Q_a\setminus Q'_a$ such that $\widehat{v}_a(Q'_a) \geq \frac{1}{19} \tau_a$ and $\widehat{v}_a(Q_a\setminus Q'_a) \geq \frac{1}{19} \tau_a$. 
				\STATE \label{step:allocA} For all $a \in A$, set $P_a = Q'_a$ 
				\STATE Set $P_n = \cup_{a \in A} \left( Q_a \setminus Q'_a \right)$ and $\sigma(a) = n $ for all $a \in A$. 
				\STATE \label{step:sigmaset} Recall that for each $a \in [n-1] \setminus A$, there exists $g' \in G \setminus P_a$ such that $v_a(g') \geq \frac{1}{19} \tau_a$. If $g' \in P_n$, then set $\sigma(a) = n$. Else, if $g' \in P_{a'}$ for $a' \in A$, then set $\sigma(a) = a'$.  
				\RETURN partition $\mathcal{P}=(P_1, P_2, \ldots, P_n)$ and mapping $\sigma$.
			\end{algorithmic}
		}
	\end{algorithm}
}

\begin{restatable}{lemma}{LemmaSubmodular}
\label{lemma:submodular}
Let  $v : 2^{{G}} \mapsto  \mathbb{R}_+$ be a monotone increasing, nonnegative, submodular valuation function over a set of indivisible goods ${G}$. If for a parameter $\kappa>0$, there exists exactly one good $\widehat{g} \in {G}$ such that $v(\{\widehat{g}\}) \geq \kappa$, then the valuation function $\widehat{v} : 2^{{G}} \mapsto  \mathbb{R}_+$, obtained using Equation (\ref{eqn:normalize}) with good $\widehat{g}$ and threshold $\kappa$, is also monotone increasing, nonnegative, and submodular.
\end{restatable}

Given fair division instance $\mathcal{I} = \langle \mathcal{A}, \mathcal{G}, \{ v_a \}_{a\in \mathcal{A}\setminus \{n\}} \  \rangle$ with a secretive agent, indivisible goods $\mathcal{G}=[m]$, and monotone, submodular valuations $v_a: 2^\mathcal{G} \mapsto \mathbb{R}_+$, write $\mu_a$ to  denote the maximin share of agent $a$ (see equation (\ref{eq:mms-defn})). In addition, let $\widehat{\mu}_a$ denote the maximin share of agent $a$ under the surrogate valuation $\widehat{v}_a$, i.e., $\widehat{\mu}_a := \max_{(A_1,\dots,A_n) \in {\Pi_n(\mathcal{G})}} \min_{j \in [n]} \widehat{v}_a(A_j)$.

Lemma \ref{lemma:mmsrelation} asserts that $\widehat{\mu}_a$ is close to $\mu_a$, for an appropriate $\kappa$.

\begin{lemma} \label{lemma:mmsrelation}
For a set of indivisible goods $G$, let $v_a : 2^{G} \mapsto  \mathbb{R}_+$ be the monotone, submodular valuation of agent $a$ and $\tau \leq \mu_a$ be a parameter such that there exists exactly one good $\widehat{g}$ of value above $\tau/19$, $v_a(\{\widehat{g}\}) \ge \tau/19$. Furthermore, let $\widehat{v} : 2^{{G}} \mapsto  \mathbb{R}_+$ be the surrogate of $v_a$ obtained using (\ref{eqn:normalize}) with good $\widehat{g}$ and $\kappa = \tau/19$. Then, we have $\widehat{\mu}_a \geq \frac{9}{19}\mu_a$. %BETTER TO STATE THE LAST INEQ IN TERMS OF \mu_a than \tau
  
%For an agent $a$ and parameter $\tau \leq $v_a : 2^{\mathcal{G}} \mapsto  \mathbb{R}_+$ be the monotone, submodular valuation of agent $a$ with the property that there exists exactly one good $g \in \mathcal{G}$ such that $v(\{g\}) \ge \tau/19$

%EXTREMELY POOR WRITING  -- TOOK ME 15 MINUTES JUST TO PARSE THROUGH THIS LEMMA STATEMENT 
%Given an agent $a$ with monotone increasing, submodular valuation function $v : 2^{\mathcal{G}} \mapsto  \mathbb{R}_+$ over the set of goods $G$ with the property that  there exists exactly one good $g \in \mathcal{G}$ such that $v(\{g\}) \ge \tau/19$, where $\tau \in \mathbb{R}_+$ is a threshold less than or equal to $\mu.$ Here, $\mu$ denotes  $n$-$\mms$ value of agent $a$ under valuation $v$. Write her normalized valuation function with respect to good $g$ as $\widehat{v} : 2^{\mathcal{G}} \mapsto  \mathbb{R}_+$ obtained by using equation \ref{eqn:normalize}, setting $\kappa$ to be $\tau/19$. Let $\widehat{\mu}$ denotes $n$-$\mms$ value of agent $a$ under valuation $\widehat{v}$. Then, $\widehat{\mu} \geq \frac{9}{19}\tau$. 
\end{lemma}

%\begin{remark2}

\noindent
{\it Remark:} The maximin share of an agent $a$ is not known a priori.\footnote{In fact, computing $\mu_a$ is an {\rm NP}-hard problem even under additive valuations--the partition problem reduces to it.} The fact that Lemma~\ref{lemma:mmsrelation} holds for $\tau$---a conservative estimate of $\mu_a$---enables us to obtain the approximation guarantee without having to compute $\mu_a$.  
%\end{remark2} 

\begin{proof} 
Write $\mathcal{P}^* = \{P^*_1, \ldots,  P^*_n\}$ to denote an $n$-partition of the indivisible goods $\mathcal{G}$ that realizes the maximin share of agent $a$; $\mu_a = \min_j v_a(P^*_j)$. From $\mathcal{P}^*$, we will create another partition $\mathcal{P}' = (P'_1, \ldots, P'_n) $ which satisfies $\widehat{v}_a(P'_i) \geq \frac{9}{19} \mu_a$ for all $i \in [n]$. This will prove the stated claim.

By reindexing, we can assume that the good $\widehat{g}$ (which uniquely satisfies $v_a( \widehat{g}) \ge \tau/19$) belongs to the first bundle $P^*_1$. Hence, via the definition of $\widehat{v}_a$, we get $\widehat{v}_a(P^*_i) = v_a(P^*_i) \geq \mu_a$ for all indices  $i \in \{2, 3, \ldots, n\}$. In addition, the lemma assumptions ensure that $\widehat{v}_a(\{ g \}) = v_a(\{g \}) < \tau/19 \leq \mu_a/19 $ for all $g \in \mathcal{G}\setminus \{ \widehat{g} \}$.

Using the property that all the goods in the bundle $P_2^* \subseteq \mathcal{G}\setminus \{ \widehat{g} \}$ have small marginals, we will next show that these exists a subset $S \subset P^*_2$ such that $\widehat{v}_a(P_1^* \cup S) \geq \frac{9}{19} \mu_a$ and $\widehat{v}_a(P_2^* \setminus S) \geq \frac{9}{19} \mu_a$. Setting $P'_1 = P_1^* \cup S$, $P'_2 = P_2^* \setminus S$, and $P'_i = P^*_i$, for all $i >2$, gives us the desired partition $\mathcal{P}' = (P'_1, \ldots, P'_n)$. 

We iteratively remove goods from $P_2^*$ (in an arbitrary order) until the total value of the removed goods exceeds $\frac{10}{19} \mu_a$. Write $S \cup \{g_\ell\}$ to denote the set of removed goods, here $g_\ell$ is the last good drawn from $P_2^*$. This procedure ensures $v_a(S \cup \{g_\ell\}) \geq \frac{10}{19} \mu_a$ and $v_a(S) < \frac{10}{19} \mu_a$. The submodularity of $v_a$ gives us $v_a(S) + v_a(\{g_\ell\}) \geq v_a(S \cup \{g_\ell\})$. Since $v_a(\{g_\ell \}) < \mu_a/19$, we get $\widehat{v}_a(S) = v_a(S) \geq \frac{9}{19} \mu_a$. The monotonicity of $\widehat{v}_a$ (Lemma~\ref{lemma:submodular}) establishes the stated bound for $P^*_1 \cup S$, i.e., $\widehat{v}_a(P^*_1 \cup S) \geq \widehat{v}_a(S) \geq \frac{9}{19} \mu_a$. 

We have an analogous bound for $P_2^* \setminus S$. Specifically, 
\begin{align*}
\widehat{v}_a (P_2^* \setminus S) & = v_a(P_2^* \setminus S) \\
& \geq v_a(P_2^*) - v_a(S) & \text{($v_a$ is submodular)} \\
& \geq \mu_a - \frac{10}{19} \mu_a = \frac{9}{19} \mu_a
\end{align*}
Therefore, the partition $\mathcal{P}' = (P^*_1 \cup S, P^*_2 \setminus S, P^*_3, \ldots, P^*_n)$ certifies that the maximin share under $\widehat{v}_a$ is at least $\frac{9}{19} \mu_a$. 
\end{proof}	

Lemma  $\ref{lemma:threshold}$ states that in Algorithm $\ref{alg:mms}$, as long as the the input threshold, $\tau_a$, for agent $a$ satisfies $\tau_a \leq \mu_a$ (and independent of the relation between $\tau_b$ and $\mu_b$, for any other agent $b$) the bundles obtained for agent $a$ approximately satisfy the $\MMS$ requirement.     

%We now prove a key property of Algorithm $\ref{alg:mms}$, which is detailed in . For an approximate secretive $\mms$ allocation, the key idea is to first create a $1/3$ $n$-$\mms$ partition of the goods and then partition each agent's bundle into two bundles of almost equal subjective value, so as to create a back-up bundle. Note that this may not always be possible under submodular valuations, due to the existence of high valued goods that makes it difficult to divide a bundle into two parts of almost equal value. We therefore resort to some appropriate preprocessing, including the normalization as detailed in equation \ref{eqn:normalize}.

\begin{lemma} \label{lemma:threshold}
For an input instance $\mathcal{I} = \langle \mathcal{A}, \mathcal{G}, \{ v_a \}_{a\in \mathcal{A}\setminus \{n\}} \  \rangle$ (with monotone, nonnegative, and submodular valuations) and thresholds $\{\tau_a\}_{a \in \mathcal{A} \setminus \{n\}}$, let  partition $\mathcal{P} = (P_1, P_2,...,P_n)$ and mapping  $\sigma:[n-1] \mapsto [n]$ be the output of Algorithm \ref{alg:mms}. If for an agent $a$ the input threshold is at most the maximin share ($\tau_a \leq \mu_a$),  then the following inequalities hold: $v_a(P_a) \geq \frac{1}{19} \tau_a$ and $v_a(P_{\sigma(a)}) \geq \frac{1}{19} \tau_a$.
%Let $\mathcal{I} = \langle \mathcal{A}, \mathcal{G}, \{ v_a \}_{a\in \mathcal{A}\setminus \{n\}} \  \rangle$ 
%Given a fair division instance of indivisible goods with a secretive agent, represented by the tuple $\mathcal{I} = \langle \mathcal{A} =[n], \mathcal{G}=[m], \{ v_a \}_{a\in \mathcal{A}\setminus \{n\}} \  \rangle$, where the valuation functions $\{ v_a \}_{a\in \mathcal{A}\setminus \{n\}}$ are monotone increasing and submodular. Let $\mathcal{P} = \{P_1, P_2,...,P_n\}$ be the allocation and $\sigma:[n-1] \mapsto [n]$ be the mapping returned by Algorithm $\ref{alg:mms}$ for this instance, with input thresholds $\tau_a \in \mathbb{R}_+$ for $a \in [n-1]$. Then, for each agent $a \in [n-1]$ whose input threshold satisfies $\tau_a \leq \mu_a$, we have $$ v_a(P_a) \geq \frac{1}{19} \tau_a, \ \text{and} \ \ v_a(P_{\sigma(a)}) \geq \frac{1}{19} \tau_a $$
\end{lemma}

%We will show that Algorithm $\ref{alg:mms}$ outputs a solution which ensures that all the agents whose input thresholds do not exceed their $n$-$\MMS$ are guaranteed to receive a $1/19$-$\MMS$ share in the presence of a secretive agent.

\begin{proof} 
	Write $A_h$ to denote the set of agents $a$ who reserve a single good of ``high value'' (i.e., $v_a(\{g_a\}) \geq \tau_a/19$) during the while-loop (Steps 2 to 4) of Algorithm $\ref{alg:mms}$. We index the agents in order they receive a bundle in the algorithm. In particular, agents in $A_h$ have a lower index than all the remaining agents in the set $[n-1] \setminus A_h$ (who are allocated a bundle in Step~\ref{step:allocA}). Note that each agent $a \in A_h$ marks one more good $g'_a$  of high value (which always exists, by the definition of $A_h$) in the remaining set of goods $G$. Therefore, for each $a \in A_h$ there exists a bundle of higher index which contains $g_a'$ and in Step~\ref{step:sigmaset} we set $\sigma(a)$ to be the index this bundle. Therefore, for each agent $a \in A_h$, we have $\sigma(a)> a$, $v_a(P_a) \geq \tau_a/19$, and $v_a(P_{\sigma(a)}) \geq \tau_a/19$; the last inequality follows from the monotonicity of $v_a$. This ensures that the required inequalities hold for each agent $a \in A_h$. The remainder of the proof shows that the output of the algorithm satisfies analogous inequalities for the remaining agents $A = \{ \mathcal{A} \setminus \{n\}\} \setminus A_h$.
	
	In particular, write $A$ to denote the set of agents left after the preprocessing performed in the while-loop (Steps 2 to 4) and $G$ be the corresponding set of remaining goods. As in Step~\ref{step:aprime}, write $A'$ to denote the set of agents $a \in A$ with exactly one high-valued good $g_a \in G$ (i.e., with exactly one good $g_a \in G$ with the property that $v_a(\{g_a\}) \geq \tau_a/19$.) Note that the algorithm considers a surrogate valuation $\widehat{v}_a$ for each $a \in A'$, and for the remaining agents (in $A\setminus A'$) the valuation remains unchanged. 
	
	With set of agents $A$, indivisible goods $G$, and valuations $\widehat{v}_a$s, the algorithm considers an $\MMS$ instance $\mathcal{I}'$ in Step~\ref{step:subinstance}, i.e., $\mathcal{I}'=\langle A, G, \{ \widehat{v}_a\}_{a \in A} \rangle$. 
	
	Note that the maximin share, $\mu^*_a$, of an agent $a \in A$ in the instance with true valuations, $\langle A, G, \{ {v}_a\}_{a \in A} \rangle$, is at least as large as $\mu_a$.\footnote{$\mu^*_a$ is obtained by maximizing over $|A|$-partitions of the set $G$, with respect to the valuation $v_a$.} Specifically, the inequality $\mu^*_a \geq \mu_a$, follows from the fact that the algorithm removes $|A_h|$ agents and exactly $|A_h|$ goods from the input instance $\mathcal{I}$. Therefore, if partition $\mathcal{B} = (B_1, \ldots, B_n)$ induces the maximin share $\mu_a$ in $\mathcal{I}$, then removing the goods allocated to agents in $A_h$ still leaves $n-|A_h| > |A|$ bundles in $\mathcal{B}$ that are contained in $G$ and are of value at least $\mu_a$. 
	
	Since $\mu_a \leq \mu^*_a$, Lemma \ref{lemma:threshold} (applied to instance $\langle A, G, \{ {v}_a\}_{a \in A} \rangle$) gives us $\widehat{\mu}_a \geq \frac{9}{19} \mu^*_a \geq \frac{9}{19} \mu_a$, here $\widehat{\mu}_a$ is the maximin share of agent $a$ in the constructed instance $\mathcal{I}'=\langle A, G, \{ \widehat{v}_a\}_{a \in A} \rangle$.\footnote{$\widehat{\mu}_a$ is obtained by maximizing over $|A|$-partitions of the set $G$, with respect to the valuation $\widehat{v}_a$.} This inequality and the $1/3$-approximation guarantee in \cite{ghodsi2018fair} ensure that if $\tau_a \leq \mu_a$, then the bundle $Q_a$ computed for agent $a$ satisfies $\widehat{v}_a(Q_a) \geq \frac{1}{3} \widehat{\mu}_a \geq \frac{3}{19} \mu_a \geq \frac{3}{19} \tau_a$. Note that the contrapositive version of this assertion gives the test for $\tau_a$ in Step~\ref{step:hightau}: if $v_a(Q_a)  < \frac{3}{19} \tau_a$, then it must be the case that $\tau_a > \mu_a$.

	Next, using the lemma assumption that $\tau_a \leq \mu_a$ and the property of the surrogate valuations, we will show that $Q_a$ can be always be partitioned into subsets $Q'_a$ and $Q_a\setminus Q'_a$ such that $\widehat{v}_a(Q'_a) \geq \tau_a/19$ and $\widehat{v}_a(Q_a\setminus Q'_a) \geq \tau_a/19$. In particular, for an agent $a \in A$, we iteratively remove goods from bundle $Q_a$ (in an arbitrary order) until the total value of the removed goods exceeds $\frac{2}{19} \tau_a$. Write $Q'_a \cup \{g_k\}$ to denote the set of removed goods, here $g_k$ is the last good drawn from $Q_a$. This procedure ensures $\widehat{v}_a(Q'_a \cup \{g_k\}) \geq \frac{2}{19} \tau_a$ and $\widehat{v}_a(Q'_a) < \frac{2}{19} \tau_a$. The submodularity of $\widehat{v}_a$ gives us $\widehat{v}_a(Q'_a) + \widehat{v}_a(\{g_k\}) \geq \widehat{v}_a(Q'_a \cup \{g_k\})$. Since the definition of surrogate function $\widehat{v}_a$ ensures that $\widehat{v}_a(\{g_k\}) \leq \tau_a/19$, the following inequality holds $\widehat{v}_a(Q'_a) \geq \frac{1}{19} \tau_a$. 
	
	In addition, via the submodularity of $\widehat{v}_a$, we have $\widehat{v}_a(Q'_a) + \widehat{v}_a(Q_a \setminus Q'_a) \geq \widehat{v}_a(Q_a) \geq \frac{3}{19}\tau_a$. That is, $\widehat{v}_a(Q_a \setminus Q'_a) \geq \frac{1}{19} \tau_a$.
	
	The fact that the algorithm successfully executes till Step~\ref{step:allocA} establishes the required property for the allocation $P_a = Q'_a$ assigned to agent $a \in A$: $v_a(P_a) \geq \widehat{v}_a (P_a) \geq \tau_a/19$. Furthermore, $v_a(P_n) \geq \widehat{v}_a (P_n) \geq \widehat{v}_a (Q_a \setminus Q'_a) \geq \tau_a/19$; here the second inequality follows from the monotonicity of $\widehat{v}_a$. Since the algorithm sets $\sigma(a) = n$ for all $a \in A$, we get the stated bounds: $\sigma(a) > a$, $v_a(P_a) \geq \frac{1}{19} \tau_a$ and $v_a(P_{\sigma(a)}) \geq \frac{1}{19} \tau_a$ for all $a \in A$. This completes the proof. 
\end{proof}

 We finally prove the main result of this section which shows that a secretive $\frac{1}{19}$-$\mms$ allocation can be computed in polynomial time.
 
\TheoremMMSsecretive*
\begin{proof}
	We will use Algorithm~\ref{alg:mms} in conjunction with a binary search to obtain the stated approximation guarantee. In particular, Lemma~\ref{lemma:threshold} ensures that Algorithm~\ref{alg:mms} successfully terminates if all the input thresholds are at most the maximin shares of the respective agents and, otherwise, it correctly identifies/flags all agents $a$ for whom $\tau_a > \mu_a$. Therefore, we can start with large enough values for the thresholds and perform a binary search to approximately find the largest possible thresholds under which Algorithm~\ref{alg:mms} succeeds in finding a partition which satisfies the required $1/19$-approximation guarantee--note that the correctness of the method (in particular, the execution of binary search) critically relies on the fact that Algorithm~\ref{alg:mms} never flags an agent $a$ with $\tau_a \leq \mu_a$. 
	
	Since Algorithm~\ref{alg:mms} runs in polynomial time, the time complexity of the overall binary search is also polynomial. Therefore, we can efficiently find a partition $\mathcal{P} = (P_1,P_2,..,P_n)$ of the indivisible goods $\mathcal{G}$ along with a mapping $\sigma$ such that for each agent $a \in \mathcal{A}\setminus \{n\}$ we have $v_a(P_a) \geq \mu_a/19 $ and $v_a(P_{\sigma(a)}) \geq \mu_a/19$. Furthermore, with our re-indexing of agents (in the order in which they receive goods), we get that $\sigma(a) > a$ for all agents $a \in \mathcal{A} \setminus \{n \}$. Therefore, using Lemma $\ref{lemma:backup}$, we can construct  $n$ bijections $\{\pi_k\}_{k \in [n]}$ which  will guarantee---for every possible choice $P_k$ of the secretive agent---to each agent a bundle of value at least $1/19$ times her maximin share. In other words, the computed partition $(P_1, \ldots, P_n)$ is a secretive $1/19$-$\MMS$ solution.
\end{proof}

% We have worked out a rudimentary constant factor by balancing certain equations, and thus the analysis in Lemma $\ref{lemma:threshold}$ can be tightened for a better approximation factor.

An improved approximation guarantee can be achieved for special subclasses of submodular functions, such as additive functions. We show that a secretive $1/2$-$\mms$ allocations for additive valuations can be computed in polynomial time in Appendix~\ref{section:additivemaximin}. \\

\noindent
{\bf Acknowledgements:} Siddharth Barman gratefully acknowledges the support of a Ramanujan Fellowship (SERB - {SB/S2/RJN-128/2015}) and a Pratiksha Trust Young Investigator Award.
\bibliographystyle{alpha}
\bibliography{references}
\appendix 
\section{Secretive 1/2-MMS Allocations Under Additive Valuations}
\label{section:additivemaximin}
This section presents a polynomial-time algorithm (Algorithm~\ref{alg:additivemms}), based on a discrete moving knife procedure, for computing secretive $1/2$-$\mms$ solutions under additive valuations.

Given a fair division instance $\langle \mathcal{A} , \mathcal{G}, \{ v_a \}_{a\in \mathcal{A}\setminus \{n\}} \  \rangle$ with a secretive agent, $m$ indivisible goods, and additive valuations.\footnote{Recall that a valuation function $v: 2^{\mathcal{G}} \mapsto \mathbb{R}^+$, over a set of indivisible goods $\mathcal{G}$, is said to be additive iff for every subset $A \subseteq G$ we have $v(A) = \sum_{g \in A}v(g)$, where $v(g)$ is the value of good $g \in G$.}  Write $\mu_a$ to denote the 
maximin share of agent $a$ over the set of goods $\mathcal{G}$; see Equation (\ref{eq:mms-defn}).

For every (non-secretive) agent $ a \in \mathcal{A}\setminus \{n\}$, we employ Woeginger's PTAS \cite{woeginger1997polynomial} to compute a threshold $\tau_a$ that satisfies $(1 - \varepsilon) \mu_a \leq \tau_a \le \mu_a$, for a small enough $\varepsilon$. We execute \cref{alg:additivemms} with these thresholds, $\tau_a$s, as input. 

{
	\begin{algorithm}
		{
			%\RaggedRight
			%\noindent
			{\bf Input:} A fair division instance $\mathcal{I} = \langle \mathcal{A} =[n], \mathcal{G}=[m], \{ v_a \}_{a\in \mathcal{A}\setminus \{n\}} \  \rangle$ with a secretive agent and additive valuations along with thresholds $\{ \tau_a \}_{a \in \mathcal{A} \setminus \{n \} }$ that satisfy $\tau_a \leq \mu_a$. \\
			%\RaggedRight{
			%\noindent
			{\bf Output:} An allocation $(P_1, \ldots, P_n)$ and a mapping $\sigma:[n-1] \mapsto [n]$ such that $v_a(P_a) \geq \tau_a/2$, $v_a(P_{\sigma(a)}) \geq \tau_a/2$, and $\sigma(a) > a$ for all $a \in \mathcal{A} \setminus \{ n \}$. 			
			\caption{Computation of $1/2$-secretive $\mms$ solution under additive valuations }	
			\label{alg:additivemms}
			\begin{algorithmic}[1]
				%\STATE Using Theorem \ref{thm:Apx-mms-value}, compute approximate maximin share $\mu'_i \in [\frac{1}{9}\mu_i,\mu_i]$ for each agent $i$.
				\STATE Initialize set of agents $A=[n-1]$ and set of goods  $G=[m]$.
				\WHILE{there exist agent $a \in A$ and good $g \in G$ such that $v_a(g) \geq  \tau_a/2$}
				\STATE \label{step:preprocess} Allocate $P_a \leftarrow\{g\}$, and update $A \leftarrow A \setminus\{a\}$ and $ G \leftarrow G\setminus\{g\}$.
				\ENDWHILE \\
				
				\WHILE{$A \neq \emptyset$}
				\STATE Arrange the goods in $G$ in an arbitrary order, $g_1, g_2, \ldots, g_{|G|}$
				\STATE \label{step:index-find} For each agent $b \in A$, let $\ell_b$ denote the minimum index such that $v_b(\{g_1, g_2, \ldots, g_{\ell_b} \}) \geq \tau_ b/2$ \\
				\COMMENT{We will prove that such an index $\ell_b < |G|$ always exists}
				\STATE \label{step:select} Select agent $a \in \argmin_{b \in A} \ell_b$.
				\STATE Assign $P_a = \{g_1, g_2, \ldots, g_{\ell_a} \}$ 
				\STATE Update $A \leftarrow A \setminus \{ a \}$ and $G \leftarrow G \setminus \{g_1, g_2, \ldots, g_{\ell_a} \}$
				\ENDWHILE
				\STATE Set $P_n = G$. 
				%\STATE For agents $a \in A$, set $\beta_a := \frac{\sum_{g \in G}v_a(g)}{|A|+1}$. Arrange the set of goods $G$ horizontally.
				%\STATE Perform a discrete moving knife procedure on the remaining goods. Each agent $a \in A$ makes the (earliest possible) cut to create a subset $S'$ such that $v_a(S' \cup g_{\ell}) \geq \beta_a$, where good $g_{\ell}$ is next good in order. Ties are broken arbitrarily.\\
				%\COMMENT {We will prove that the above step creates a partition $\mathcal{Q} = (Q_1, Q_2, \ldots, Q_{|A|+1})$ for the reduced instance}.
				%\STATE For each $a \in A$, set $P_a = Q_a$ and set $P_n = Q_{|A|+1}$. Index the bundles in the order of their creation.
				\STATE \label{step:sigma-find} Index the agents in order they receive a bundle in the algorithm and for each agent $a \in [n-1]$ find a bundle $P_{a'}$ such that $a'>a$ and $v_a(P_{a'}) \geq \tau_a/2$. Set $\sigma(a) = a'$. \\
				\COMMENT{We will prove that such a higher-index bundle exists for each agent}
				\RETURN partition $\mathcal{P}=(P_1, P_2, \ldots, P_n)$ and mapping $\sigma$.
			\end{algorithmic}
		}
	\end{algorithm}
}

\begin{restatable}{theorem}{Theorem1/2MMSsecretive}
	\label{theorem:1/2mmssecretive}
	For every fair division instance $\mathcal{I} = \langle \mathcal{A}, \mathcal{G}, \{ v_a \}_{a\in \mathcal{A}\setminus \{n\}} \  \rangle$ with a secretive agent and additive valuations, a secretive $\frac{1}{2}$-$\mms$ solution exists and can be computed in polynomial time.
\end{restatable}

\begin{proof}
	As in \cref{alg:additivemms}, we index the agents in the same order in which they receive a bundle in the algorithm. Write $A_h$ to denote the set of agents that receive a single good of high value (i.e., of value at least $\tau_a/2$) in the preprocessing round (Step~\ref{step:preprocess}) of the algorithm.  Denote the set of goods assigned to agents in this round as $G_h$. Note that $|A_h| = |G_h|$ and let $r := n - 1 - |A_h|$ denote the number of remaining non-secretive agents that receive a bundle during the second while-loop of the algorithm.  The above-mentioned reindexing ensures that $A_h =\{1, 2, \ldots, n-1-r\}$.

	Note that for any agent $a \in \mathcal{A} \setminus \{n \}$, the maximin share restricted to $(r+1)$ agents and the set of goods $\mathcal{G} \setminus G_h$ is at least as large as $\mu_a$.  Specifically, if maximin share $\mu_a^*$ is obtained by maximizing over $(r+1)$-partitions of the set $\mathcal{G} \setminus G_h$, with respect to the valuation $v_a$, then $\mu^*_a \geq \mu_a$. This follows from considering an $n$-partition $(B_1, B_2, \ldots, B_n)$ of $\mathcal{G}$ that induces the maximin share, i.e., $v_a(B_i) \geq \mu_a$ for all $ i \in [n]$. Since we are removing $n-r-1 = |G_h|$ goods and an equal number of agents from consideration, there will exist $(r+1)$ bundles, $B_i$s, that are contained in $ \mathcal{G} \setminus G_h$. In order words, $ \mathcal{G} \setminus G_h$ admits an $(r+1)$-partition in which each bundle is of value at least $\mu_a$, i.e., we have $\mu^*_a \geq \mu_a$. Recall that the input threshold $\tau_a$ is at most the maximin share, hence $\mu^*_a \geq \tau_a$. 
	
	Next, we will show that Steps~\ref{step:index-find} and~\ref{step:sigma-find} in the algorithm always succeed. This in turn implies that the algorithm returns a partition $\mathcal{P} = (P_1, \ldots, P_n)$ along with a mapping $\sigma$, which satisfies the conditions in~\cref{lemma:backup}. Hence, we get that \cref{alg:additivemms} finds a secretive $1/2$-$\MMS$ solution.

	Since the valuations of all the agents $a \in [n-1]$ are additive, we have $\frac{v_a(\mathcal{G} \setminus G_h)}{r+1} \geq \mu^*_a \geq \tau_a$. That is, for every agent $a$ the following bound holds $v_a(\mathcal{G} \setminus G_h) \geq (r+1) \tau_a$. Hence, at the beginning of the second while-loop (in Step~\ref{step:index-find}), for every agent $a \in [n-1] \setminus A_h$, there exists an index $\ell_a$ such that $v_a(\{g_1, g_2, \ldots, g_{\ell_a} \}) \geq \tau_a/2$. 
	
	We will inductively prove that this property continues to hold, i.e., in any iteration of the second while-loop, if $A$ and $G$ are the set of remaining agents and goods, respectively, then for all $a \in A$ we have $v_a(G) \geq \tau_a$. This will imply that in Step~\ref{step:index-find} for every remaining agent $b \in A$ we have a well-defined index $\ell_b$. Let $P_{(1)}, P_{(2)}, \ldots, P_{(t)}$ be the set of bundles allocated in the first $t$ iterations of the second while-loop. For a remaining agent $a \in A$, we have $v_a(P_{(s)}) \leq \tau_a$, for $1 \leq s \leq t$. This follows, from the fact that, for agent $a$, a (strict) subset of $P_{(s)}$ with one less good, say $P_{(s)} \setminus \{g_\ell\}$, was of value strictly less than $\tau_a/2$ (else, $a$ would have been selected in Step~\ref{step:select}). The preprocessing performed in the algorithm ensures that, for all agents in $A$ and goods $g \in G$, we have $v_a(g) \leq \tau_a/2$. Hence, $v_a( P_{(s)}) = v_a( P_{(s)}\setminus \{g_\ell\} ) + v_a ( \{g_\ell\}) \leq \tau_a/2 + \tau_a/2$. Recall that $v_a( \mathcal{G} \setminus G_h) \geq (r+1) \tau_a$. Therefore, in iteration $t \leq r$, for the set of remaining goods, $G = \mathcal{G} \setminus \left( G_h \cup  (\cup_{s=1}^{t} P_{(s)}) \right)$, the required bound holds, $v_a(G) \geq  (r+1 - t) \tau_a$. 
	
	%Note that, this argument, applied with $t=r$, shows that for the $(n-1)$th agent (by the employed indexing, this is the last agent to be processed by the algorithm) the set of goods $P_n$ are of value $\tau_a$, i.e., $v_{n-1}(P_n) \geq (r+1 - r) \tau_a = \tau_a$. 
	
	These observations ensure that the algorithm successfully associates with each agent $a \in [n-1]$ a bundle $P_a$ with the property that $v_a(P_a) \geq \tau_a/2$. 
	
	To complete the proof we will show that the algorithm also succeeds in Step~\ref{step:sigma-find}. As mentioned previously, for every agent $a \in [n-1]$ we have $\mu^*_a \geq \mu_a$. Consider an agent $a \in A_h$ and note that for $a$'s additive valuation we have $\frac{v_a(\mathcal{G} \setminus G_h)}{r+1} \geq \mu^*_a$.  The set of goods $\mathcal{G} \setminus G_h $ gets partitioned into $(r+1)$ bundles $P_{n-r}, \ldots, P_n$ that satisfy ${\sum_{s=n-r}^n v_a(P_s)} = v_a(\mathcal{G} \setminus G_h)$. By an averaging argument, we get that there exists an index $a' \geq n-r > a$ such that $v_a(P_{a'}) \geq \mu^*_a \geq \tau_a$. That is, for all $a \in A_h$, the $a'$ required in Step~\ref{step:sigma-find} exists.\footnote{Of course, such an $a'$ can be found efficiently by enumeration.} 
	
	In the complementary case we consider agents $a \notin A_h$. Let $t$ denote the iteration (of the second while-loop) in which agent $a$ is processed. If $G$ is the set of goods that remain unassigned at the beginning of iteration $t$, then (via the arguments mentioned previously) we have $v_a(G) \geq (r+1 - t-1)  \tau_a$. Furthermore, note that $v_a(P_a) \leq \tau_a$; again, this is a consequence of preprocessing away the high-valued goods.  The set of goods left after processing $a$ is $G \setminus P_a$ and for $a$ the value of these goods $v_a( G \setminus P_a) \geq (r+1 -t) \tau_a$. The algorithm creates $r+1-t$ bundles from these goods (including $P_n$). Therefore, for $a$, one of these later-created bundles is guaranteed to be of value at least $\tau_a$.  Hence, for agents $a \in [n-1] \setminus A_h$ as well, we get that the index $a'$---required to construct mapping $\sigma$ in Step~\ref{step:sigma-find}---always exists.  
	
	Overall, by applying Lemma $\ref{lemma:backup}$ on the mapping $\sigma$, we can construct the appropriate bijections that prove that the returned partition $\mathcal{P}$ is a secretive $1/2$-$\mms$ solution. This completes the proof. 
	
\end{proof}	

\noindent
{\it Remark:} It is possible to bypass the application of Woeginger's PTAS by employing a binary-search method (similar to the one used in Section~\ref{section:mms}). However, we have used the PTAS here for ease of exposition and analysis. 

\section{Proof of \cref{lemma:submodular}} %Preservation of Submodularity and Monotonicity}
\label{section:submodularmono}
In this section we prove that transformation detailed in Equation $\ref{eqn:normalize}$ preserves submodularity and monotonicity.

\LemmaSubmodular*

\begin{proof}
Write $v$ to denote the initial submodular function and $\widehat{v}$ be the transformed function. We will prove that $\widehat{v}$ is also submodular; in particular, we will establish that for any pair of subsets, $A, B \subseteq G$, the following inequality holds $\widehat{v}(A) + \widehat{v}(B) \geq \widehat{v}(A \cup B) + \widehat{v}(A \cap B)$.  The proof relies on analyzing the following cases, which depend upon the containment of the good $\widehat{g}$, with respect to the sets $A$ and $B$. \\
	
	\noindent
	\emph{Case {\rm I}:} Good $\widehat{g} \notin A \cup B$ 
	\begin{align*}
		\widehat{v}(A) + \widehat{v}(B) & = v(A) + v(B) & \text{(since, $\widehat{g} \notin A \cup B$)} \\
		& \geq v(A \cup B) + v(A \cap B) & \text{(by submodularity of $v$)}\\
		& = \widehat{v}(A \cup B) + \widehat{v}(A \cap B) & \text{(since $\widehat{g} \notin A \cup B$)} 
	\end{align*} 
	
	\noindent
	\emph{Case {\rm II}:} Good $\widehat{g} \in A \setminus B$
	\begin{align*}
		\widehat{v}(A) + \widehat{v}(B) & = v(A\setminus\{ \widehat{g}\}) + \min\left\{ \kappa,   v_{A\setminus \{\widehat{g} \}} (\widehat{g})  \right\} + {v}(B) & \text{(since $\widehat{g} \in A$ and $\widehat{g} \notin B$)}\\
		& \geq v(A\setminus\{ \widehat{g}\}) + \min\left\{ \kappa,   v_{(A \cup B)\setminus \{\widehat{g} \}} (\widehat{g})  \right\} + {v}(B) & \text{(by submodularity of $v$)}\\
		& \geq v((A \cup B) \setminus \{\widehat{g}\}) + v(A \cap B) +  \min\left\{ \kappa,   v_{(A \cup B)\setminus \{\widehat{g} \}} (\widehat{g})  \right\} & \text{(by submodularity of $v$)}\\
		& = v((A \cup B) \setminus \{\widehat{g}\}) +  \min\left\{ \kappa,   v_{(A \cup B)\setminus \{\widehat{g} \}} (\widehat{g})  \right\} + v(A \cap B) & \\
		& = \widehat{v}(A \cup B) + {v}(A \cap B) & \text{(by the definition of $\widehat{v}$)}\\
		& = \widehat{v}(A \cup B) + \widehat{v}(A \cap B) & \text{($\widehat{g} \notin A \cap B$)} 
	\end{align*}
	
	\noindent
	\emph{Case {\rm III}:} Good $\widehat{g} \in B \setminus A$: Identical to \emph{Case {\rm II}} \\
	
	\noindent
	\emph{Case {\rm IV}:} Good $\widehat{g} \in A \cap B$\\
	There are four subcases based on the value of $\kappa$ relative to $v_{A\setminus\{ \widehat{g}\}}(\widehat{g}) $ and $v_{B\setminus\{ \widehat{g}\}}(\widehat{g})$.\\ 
	\noindent	\emph{Subcase (i):} $v_{A\setminus\{ \widehat{g}\}}(\widehat{g}), \ v_{B\setminus\{ \widehat{g}\}}(\widehat{g}) \leq \kappa$
	\begin{align*}
		\widehat{v}(A) + \widehat{v}(B) & = v(A\setminus\{ \widehat{g}\}) + \min\left\{ \kappa,   v_{A\setminus \{\widehat{g} \}} (\widehat{g})  \right\} +v(B\setminus\{ \widehat{g}\}) + \min\left\{ \kappa,   v_{B\setminus \{\widehat{g} \}} (\widehat{g}) \right\} \qquad \text{(by definition of $\widehat{v}$)}\\
		& = v(A\setminus\{ \widehat{g}\}) + v_{A\setminus \{\widehat{g} \}} (\widehat{g}) + v(B\setminus\{ \widehat{g}\}) + v_{B\setminus \{\widehat{g} \}} (\widehat{g})  \qquad \text{(since $v_{A\setminus\{ \widehat{g}\}}(\widehat{g}), \ v_{B\setminus\{ \widehat{g}\}}(\widehat{g}) \leq \kappa$)}\\
		& = v(A) + v(B) \\
		& \geq v(A \cup B) + v(A \cap B)  \qquad \qquad \qquad \qquad \qquad \qquad \qquad \text{(by submodularity of $v$)}\\
		& = v((A \cup B) \setminus \{\widehat{g}\}) +  v_{(A \cup B)\setminus \{\widehat{g} \}} (\widehat{g}) +v((A \cap B) \setminus \{\widehat{g}\}) +  v_{(A \cap B)\setminus \{\widehat{g} \}} (\widehat{g}) \\
		& \geq v((A \cup B) \setminus \{\widehat{g}\}) +   \min\left\{ \kappa,  v_{(A \cup B)\setminus \{\widehat{g} \}} (\widehat{g}) \right\} +v((A \cap B) \setminus \{\widehat{g}\})   +   \min\left\{ \kappa, v_{(A \cap B)\setminus \{\widehat{g} \}} (\widehat{g}) \right\}\\
		& = \widehat{v}(A \cup B) + \widehat{v}(A \cap B) \qquad \qquad \qquad \qquad \qquad \qquad \qquad  \text{(since $\widehat{g} \in A$ and $\widehat{g} \in B$)}
	\end{align*}
	
	\indent	\emph{Subcase (ii):} $v_{A\setminus\{ \widehat{g}\}}(\widehat{g}) \geq \kappa$ and $  v_{B\setminus\{ \widehat{g}\}}(\widehat{g}) \leq \kappa$
	\begin{align*}
		\widehat{v}(A) + \widehat{v}(B) & = v(A\setminus\{ \widehat{g}\}) + \min\left\{ \kappa,   v_{A\setminus \{\widehat{g} \}} (\widehat{g})  \right\} +v(B\setminus\{ \widehat{g}\}) + \min\left\{ \kappa,   v_{B\setminus \{\widehat{g} \}} (\widehat{g})  \right\}  \tag{since $\widehat{g} \in A$ and $\widehat{g} \in B$}\\
		& = v(A\setminus\{ \widehat{g}\}) + \min\left\{ \kappa,   v_{A\setminus \{\widehat{g} \}} (\widehat{g})  \right\}  + v(B\setminus\{ \widehat{g}\}) + v_{B\setminus \{\widehat{g} \}} (\widehat{g})  \tag{since, $v_{B\setminus\{ \widehat{g}\}}(\widehat{g}) \leq \kappa$} \\
		%& \geq v(A\setminus\{ g\}) + \min\left\{ \kappa,   v_{A\setminus \{g \}} (g)  \right\}  + v(B\setminus\{ g\}) + v_{B\setminus \{g \}} (g) \qquad \qquad \qquad \qquad  \ \  \text{(since, $v_{B\setminus\{ g\}}(g) \leq \kappa$)}\\
		& \geq v((A \cup B) \setminus \{\widehat{g}\}) + v((A \cap B) \setminus \{\widehat{g}\}) + \min\left\{ \kappa,   v_{A\setminus \{\widehat{g} \}} (\widehat{g})  \right\}+ v_{B\setminus \{\widehat{g} \}} (\widehat{g})   \tag{by submodularity of $v$} \\
		& = v((A \cup B) \setminus \{\widehat{g}\}) + v((A \cap B) \setminus \{\widehat{g}\}) + \kappa + v_{B\setminus \{\widehat{g} \}} (\widehat{g}) \tag{since $v_{A\setminus \{\widehat{g} \}} (\widehat{g})  \geq \kappa$} \\
		& = v((A \cup B) \setminus \{\widehat{g}\}) + v((A \cap B) \setminus \{\widehat{g}\}) + \min\left\{ \kappa,   v_{(A \cap B) \setminus \{\widehat{g} \}} (\widehat{g})  \right\} + v_{B\setminus \{\widehat{g} \}} (\widehat{g})  \tag{since $v_{(A \cap B) \setminus \{\widehat{g} \}} (\widehat{g}) \geq v_{A\setminus \{\widehat{g} \}} (\widehat{g}) \geq \kappa $} \\
		% v_{(A \cap B) \setminus \{\widehat{g} \}} (\widehat{g}) \\ %\text{(since $v_{(A \cap B) \setminus \{\widehat{g} \}} \ge v_{A \setminus \{\widehat{g}}(\widehat{g}) \ge \kappa$)}\\
		& \geq v((A \cup B) \setminus \{\widehat{g}\}) + v((A \cap B) \setminus \{\widehat{g}\})   +   \min\left\{ \kappa, v_{(A \cap B)\setminus \{\widehat{g} \}} (\widehat{g}) \right\} +  \min\left\{ \kappa,  v_{(A \cup B)\setminus \{\widehat{g} \}} (\widehat{g}) \right\} \tag{submodularity of $v$} \\
		& = \widehat{v}(A \cup B) + \widehat{v}(A \cap B)  \tag{since $\widehat{g} \in A$ and $\widehat{g} \in B$}
	\end{align*}
	
	\indent	\emph{Subcase (iii):} $v_{A\setminus\{ \widehat{g}\}}(\widehat{g}) \leq \kappa$ and $  v_{B\setminus\{ \widehat{g}\}}(\widehat{g}) \geq \kappa$: Identical to \emph{Subcase (ii)} \\
	
	\indent	\emph{Subcase (iv):} $v_{A\setminus\{ \widehat{g}\}}(\widehat{g}),  v_{B\setminus\{ \widehat{g}\}}(\widehat{g}) < \kappa$\\
	
	\begin{align*}
		\widehat{v}(A) + \widehat{v}(B) & = v(A\setminus\{ \widehat{g}\}) + \min\left\{ \kappa,   v_{A\setminus \{\widehat{g} \}} (\widehat{g})  \right\} +v(B\setminus\{ \widehat{g}\}) + \min\left\{ \kappa,   v_{B\setminus \{\widehat{g} \}} (\widehat{g})  \right\}  \qquad \quad \text{(since $g \in A$ and $g \in B$)}\\	
		%& = v(A\setminus\{ g\}) + \min\left\{ \kappa,   v_{(A \cap B) \setminus \{g \}} (g)  \right\}  + v(B\setminus\{ g\}) + v_{B\setminus \{g \}} (g) \qquad \qquad \quad \ \ \text{(by submodularity of $v$)}\\
		%& \geq v(A\setminus\{ g\}) + v(B\setminus\{ g\}) + \min\left\{ \kappa,   v_{(A \cap B) \setminus \{g \}} (g)  \right\} + v_{(A \cup B) \setminus \{g \}} (g)  \\
		%& \geq v(A \cup B)\setminus\{ g\}) + v((A \cap B) \setminus\{ g\}) + \min\left\{ \kappa,   v_{(A \cap B) \setminus \{g \}} (g)  \right\} + v_{(A \cup B) \setminus \{g \}} (g) \\
		& = v(A\setminus\{ \widehat{g} \}) +  v_{(A) \setminus \{\widehat{g} \}} (\widehat{g})  + v(B\setminus\{ \widehat{g}\}) + v_{B\setminus \{\widehat{g} \}} (\widehat{g})\\
		& = v(A) + v(B) \\
		& \ge v(A \cup B) + v(A \cap B) \qquad \qquad \qquad \qquad \qquad \qquad \qquad \qquad \qquad\qquad \qquad \text{(by submodularity of $v$)} \\
		& =v((A \cup B) \setminus \{\widehat{g}\}) +   v_{(A \cup B)\setminus \{\widehat{g} \}} (\widehat{g}) +v((A \cap B) \setminus \{\widehat{g}\})   +   v_{(A \cap B)\setminus \{\widehat{g} \}} (\widehat{g}) \\
		& = v((A \cup B) \setminus \{\widehat{g}\}) +   \min\left\{ \kappa,  v_{(A \cup B)\setminus \{\widehat{g} \}} (\widehat{g}) \right\} +v((A \cap B) \setminus \{\widehat{g}\})   +   \min\left\{ \kappa, v_{(A \cap B)\setminus \{\widehat{g} \}} (\widehat{g})\right\} \\
		& = \widehat{v}(A \cup B) + \widehat{v}(A \cap B)  \qquad \qquad \qquad \qquad \qquad \qquad \qquad \qquad \qquad \qquad \qquad \ \  \text{(by definition of $\widehat{v}$)}
	\end{align*}
	This shows that submodularity is preserved when we transform $v$ to $\widehat{v}$.

	For monotonicity, we will show that $\widehat{v}(A \cup \{h\}) \geq \widehat{v}(A)$ for all subsets $A \subseteq G$ and for all goods $h \in G$. Note that the claim is trivial when $h \in A$, since in this case $A \cup \{ h \} = A$.  Therefore, for the remainder of the proof we will focus on cases wherein $h \notin A$.
	
	%Note that since $v$ is monotone, $v_A(h)  \geq0$  for all subsets $A$ and goods $h \in G$. Also, $v(h) \geq 0$ for all goods $h \in G$ and $\kappa >0$. This implies that $\widehat{v}_A(h) \geq 0$, and thus completing the proof. 

	We have four cases to consider based on the containment of good $\widehat{g}$ in $A$ and whether or not the good $h$ is the same as $\widehat{g}$.
	
	\noindent
	\emph{Case {\rm I}:} Good $\widehat{g} \notin A$ and  $\widehat{g} \neq h$. 
	
	\begin{align*}
		\widehat{v}(A \cup \{h\}) &= v(A \cup \{h\})  & \text{(since $\widehat{g} \notin A \cup \{ h \}$ )} \\
		& \geq v(A) &  \text{(by monotonicity of $v$)}\\
		& = \widehat{v}(A) & \text{(since $\widehat{g} \notin A$ )}
	\end{align*} 	
	
	\noindent
	\emph{Case {\rm II}:} Good $\widehat{g} \in A$ and  $\widehat{g} = h$. \\
	Note that in this case $h \in A$ and, hence, the required inequality holds directly. \\
	
	\noindent
	\emph{Case {\rm III}:} Good $\widehat{g} \notin A$ and  $\widehat{g} = h$. \\
	Here, the required inequality reduces to $\widehat{v}(A \cup \{\widehat{g}\}) \geq \widehat{v}(A)$. Since $\widehat{g} \notin A$, we have 
	
	\begin{align*}
		\widehat{v}(A \cup \{\widehat{g}\}) &= v(A) + \min\left\{ \kappa, v_A(\widehat{g}) \right\} & \text{(by definition of $\widehat{v}$)} \\
		& \geq v(A) &  \text{(since, $\min\left\{ \kappa, v_A(\widehat{g}) \right\} \geq 0$ )}\\
		& = \widehat{v}(A) & \text{(since $\widehat{g} \notin A$ )}
	\end{align*}
	
	\noindent
	\emph{Case {\rm IV}:} Good $\widehat{g} \in A$ and  $\widehat{g} \neq h$. \\
	In this case, first we express  $\widehat{v}(A \cup \{h\})$ and $\widehat{v}(A)$ using the definition of $\widehat{v}$:
	\begin{align*}
		\widehat{v}(A \cup \{h\}) &=v((A \setminus \{\widehat{g}\}) \cup \{h\}) + \min\left\{\kappa, v_{(A \setminus \{\widehat{g} \}) \cup \{h\}} (\widehat{g})\right\} 
	\end{align*}
	\begin{align*}
		\widehat{v}(A) & =v(A \setminus \{\widehat{g}\}) + \min\left\{\kappa, v_{A \setminus \{\widehat{g}\}} (\widehat{g})\right\}
	\end{align*}
	\indent	\emph{Subcase (i):} $v_{(A \setminus \{\widehat{g} \}) \cup \{h\}} (\widehat{g}) \ge \kappa$
		\begin{align*}
		\widehat{v}(A \cup \{h\}) &= v((A \setminus \{\widehat{g}\}) \cup \{h\}) + \min\left\{\kappa, v_{(A \setminus \{\widehat{g} \}) \cup \{h\}} (\widehat{g})\right\}\\
		&= v((A \setminus \{\widehat{g}\}) \cup \{h\}) + \kappa \\
		&\ge v(A \setminus \{\widehat{g}\}) + \kappa \tag{by monotonicity of $v$}\\
		&=  v(A \setminus \{\widehat{g}\}) + \min \left \{\kappa, v_{A \setminus \{\widehat{g}\}} (\widehat{g})\right\} \tag{$v_{A \setminus \{\widehat{g}\}} (\widehat{g}) \geq v_{(A \setminus \{\widehat{g} \}) \cup \{h\}} (\widehat{g}) \geq \kappa$} \\
		&= \widehat{v}(A)
	\end{align*}
	
	\indent \emph{Subcase (ii):} $v_{(A \setminus \{\widehat{g} \}) \cup \{h\}} (\widehat{g}) < \kappa$ 
	\begin{align*}
		\widehat{v}(A \cup \{h\}) &= v((A \setminus \{\widehat{g}\}) \cup \{h\}) + \min\left\{\kappa, v_{(A \setminus \{\widehat{g} \}) \cup \{h\}} (\widehat{g})\right\}\\
		&= v((A \setminus \{\widehat{g}\}) \cup \{h\}) + v_{(A \setminus \{\widehat{g} \}) \cup \{h\}} (\widehat{g})\\
		&= v(A \cup \{h\}) \\
		&\ge v(A) \qquad \qquad \qquad\qquad\qquad \qquad \qquad \qquad \qquad \qquad \text{(by monotonicity of $v$)}\\
		&= v(A \setminus \{\widehat{g}\}) + v_{A \setminus \{\widehat{g}\}}(\widehat{g})\\
		&\ge v(A \setminus \{\widehat{g}\}) + \min \left \{\kappa, v_{A \setminus \{\widehat{g}\}}(\widehat{g}) \right \}\\
		&= \widehat{v}(A)
	\end{align*}
\end{proof}

\end{document}